\documentclass[preprint2]{aastex}

\def\bd28{BD\,+28$^{\circ}$\,4211}
\def\gvel{$\gamma^{2}$\,Vel}
\def\zpup{$\zeta$\,Pup}
\def\dori{$\delta$\,Ori~A}
\def\fuse{{\it FUSE\/}}

\def\hst{{\it HST\/}}
\def\oao{{\it Copernicus}}
\def\orf{{\it ORFEUS-SPAS II}}
\def\lya{Ly$\alpha$}
\def\lyb{Ly$\beta$}
\def\lyg{Ly$\gamma$}
\def\lyd{Ly$\delta$}
\def\lye{Ly$\epsilon$}
\def\lyz{Ly$\zeta$}
\def\lyeta{Ly$\eta$}
\def\lyth{Ly$\theta$}
\def\lyi{Ly$\iota$}

\def\EE#1{\times 10^{\small#1}}
\def\cmsq{\rm ~cm^{\small -2}}
\def\cm3{\rm ~cm^{\small -3}}
\def\kms{\rm ~km~s^{\small -1}}

\def\ergcmsA{\rm\,erg~cm^{\small -2}~s^{\small -1}~\AA^{\small -1}}

\def\wl{$\lambda$}
\def\wll{$\lambda\lambda$}
\def\h2{H$_2$}
\def\hi{{\ion{H}{1}\,}}

\def\di{{\ion{D}{1}\,}}
\def\nh2{$N(\rm H_2)$\,}
\def\nhi{$N$({\ion{H}{1})\,}}
\def\ndi{$N$({\ion{D}{1})\,}}
\def\nni{$N$({\ion{N}{1})\,}}
\def\nnii{$N$({\ion{N}{2})\,}}
\def\noi{$N$({\ion{O}{1})\,}}

\def\heii{{\ion{He}{2}}}

\def\ni{{\ion{N}{1}}}
\def\nii{{\ion{N}{2}}}

\def\oi{{\ion{O}{1}}}

\def\svi{{\ion{S}{6}}}
\def\pii{{\ion{P}{2}}}
\def\nai{{\ion{Na}{1}}}

\def\feii{{\ion{Fe}{2}}}
\def\chisq{$\chi^2$}
\def\chisqmin{$\chi^{2}_{min}$}

\def\teff{$T_{eff}$}
\def\eqw{$W_{\lambda}$}

\slugcomment{Accepted for publication in the Astrophysical Journal}
\shortauthors{SONNEBORN ET AL.}
\shorttitle{DEUTERIUM TOWARD \bd28}

\begin{document}
\title{Interstellar Deuterium, Nitrogen, and Oxygen Abundances Toward \bd28:
 Results from the Far Ultraviolet Spectroscopic Explorer\footnote{Based on observations made with the NASA-CNES-CSA Far Ultraviolet Spectroscopic Explorer. {\it FUSE} is operated for NASA by the Johns Hopkins University under NASA contract NAS5-32985.}}

\author{George Sonneborn,\altaffilmark{2}
Martial Andr\'e,\altaffilmark{3}
Cristina Oliveira,\altaffilmark{3}
Guillaume H\'ebrard,\altaffilmark{4}\\
J. Christopher Howk,\altaffilmark{3}
Todd M. Tripp,\altaffilmark{5}
Pierre Chayer,\altaffilmark{3,6}
Scott D. Friedman,\altaffilmark{3}\\
Jeffrey W. Kruk,\altaffilmark{3}
Edward B. Jenkins,\altaffilmark{5}
Martin Lemoine,\altaffilmark{4}
H. Warren Moos,\altaffilmark{3}\\
William R. Oegerle,\altaffilmark{2}
Kenneth R. Sembach,\altaffilmark{3} and
Alfred Vidal-Madjar\altaffilmark{4}
}
\affil{}

\altaffiltext{2}{Laboratory for Astronomy and Solar Physics, Code 681,
NASA Goddard Space Flight Center, Greenbelt, MD 20771; 
~george.sonneborn@gsfc.nasa.gov, oegerle@s2.gsfc.nasa.gov}
\altaffiltext{3}{Dept. of Physics and Astronomy, The Johns Hopkins University, Baltimore, MD  21218; ~mandre, oliveira, howk, chayer, scott, kruk, hwm, and sembach @pha.jhu.edu.}
\altaffiltext{5}{Princeton University Observatory, Princeton, NJ 08544;
~tripp and ebj@astro.princeton.edu.}
\altaffiltext{4}{Institut d'Astrophysique de Paris, 98bis Blvd. 
Arago, 75014 Paris, France;  alfred, hebrard, and lemoine@iap.fr.}
\altaffiltext{6}{Primary affiliation: Department of Physics and Astronomy, University of Victoria, P. O. Box 3055, Victoria, BC V8W 3P6, Canada.}

\begin{abstract}

High resolution far-ultraviolet spectra of the O-type subdwarf \bd28\ were obtained with the Far Ultraviolet Spectroscopic Explorer to measure the interstellar deuterium, nitrogen, and oxygen abundances in this direction. The interstellar \di\ transitions are analyzed down to \lyi\ at 920.7 \AA. The star was observed several times at different target offsets in the direction of spectral dispersion. The aligned and coadded spectra  have high signal-to-noise ratios (S/N $=50-100$).  \di, \ni, and \oi\  transitions were analyzed with curve-of-growth and profile fitting techniques.  A model of interstellar molecular hydrogen on the line of sight was derived from \h2\ lines in the \fuse\ spectra and used to help analyze some features where blending with \h2\ was significant.  The \hi\ column density was determined from high resolution \hst /STIS spectra of \lya\ to be log \nhi\ $=19.846 \pm 0.035 ~(2\sigma)$, which is higher than is typical for sight lines in the local ISM studied for D/H.  We found that D/H $=(1.39 \pm 0.21) \EE{-5} ~(2\sigma)$ and O/H $=(2.37 \pm 0.55) \EE{-4} ~(2\sigma)$.  O/H toward \bd28\ appears to be significantly below the mean O/H ratio for the ISM and the Local Bubble.  

\end{abstract}


\section{INTRODUCTION}

The abundance ratio of atomic deuterium to hydrogen (D/H) is a key diagnostic of light element production and the baryon-to-photon ratio $(\eta)$ in big bang nucleosynthesis (BBN, Walker et al. 1991).   Significant efforts in recent years to measure D/H in  high redshift intergalactic gas have generally yielded  values  in the range D/H $=(2.5-4.0) \EE{-5}$ (O'Meara et al. 2001 and references therein).

In galaxies, D/H is an important indicator of chemical evolution (Audouze \& Tinsley 1974; Boesgard \& Steigman 1985; Tosi et al. 1998), as D is easily destroyed in stars ($^{2}{\rm H} + p \rightarrow ^{3}{\rm He} + \gamma$) at temperatures of only a few million degrees. In fact, there are no known appreciable sources of D other than BBN.  The interstellar medium (ISM), as the repository of stellar mass loss, holds the keys to unravelling the chemical history of D over the age of the galaxy.  Measurements of atomic D/H in the ISM were first made by the \oao\ satellite toward bright OB stars within $\sim500$ pc of the sun (Rogerson \& York 1973; see Vidal-Madjar \& Gry 1984 for a review).  Significant progress had been made using the Hubble Space Telescope (\hst) to obtain some precise D/H measurements in the local ISM (Linsky et al. 1998; Vidal-Madjar et al. 1998), but these have necessarily been limited to very low column density lines of sight (\nhi$\sim 10^{18} \cmsq$) where \lya\ can be used to measure the \di\ column density.

The D/H ratio was measured at high spectral resolution ($\lambda/\Delta\lambda\sim 80,000$) with  the Interstellar Medium Absorption Profile Spectrograph (IMAPS, Jenkins et al. 1988, 1999)  toward three OB stars (\dori, Jenkins et al. 1999; \gvel\ and \zpup, Sonneborn et al. 2000) first studied by \oao. These studies showed unequivocally that there is a large spread in D/H (factor of 3) on sight lines toward early type stars at distances of 300-500 pc. The IMAPS instrument was limited to observing very bright OB stars by its instrument design.  At that time IMAPS could only observe the lower Lyman series through \lye\ \wl937 because its gratings were coated with lithium-fluoride (LiF). 

\fuse\ provides a new capability to study the entire Lyman series (except \lya) in much fainter stars than were possible with \oao\ or IMAPS, and hence D/H in the ISM within 200 pc can now be studied systematically for the first time.  This is important because the lines of sight to nearby stars are usually simpler with less blending of different velocity components.  \fuse\ has sufficient sensitivity to perform absorption line studies of white dwarfs as well as extragalactic objects. Both categories of targets were inaccessible to \oao\ or IMAPS.  The study of D/H throughout the Galaxy is one of the primary objectives of the \fuse\ mission (Moos et al. 2000).

In this paper we report the analysis of D, N, and O abundances toward the O subdwarf \bd28\ using high signal-to-noise (S/N) ratio spectra obtained with \fuse.  \bd28\ has a very high temperature ($\sim$82,000 K), lies at a distance of $\sim100$ pc, and has a \hi\ column density, \nhi, of $\sim10^{20} \cmsq$. This column density is much higher than those of the other LISM lines of sight studied previously for D/H, and very similar to that of the D/H sight lines studied by IMAPS.  In this \nhi\ regime the Lyman series shortward of \lyd\ \wl949 is required to study \ndi.

Because a measurement of D/H requires both accurate \ndi\ and \nhi,  we use \fuse\ spectra to measure total column densities for \ndi, \nni, and \noi and \hst/STIS spectra of \lya\ to measure \nhi.  Even in the regime \nhi$\sim 10^{20} \cmsq$, the damping wings of \lyb\ and higher Lyman series transitions are too weak to obtain an accurate \nhi\ measurement.

The observations and data processing are presented in \S 2 and the stellar atmosphere model and synthetic spectrum in \S 3.  The measurement of \nhi\ is discussed in \S 4. In \S 5 we describe the measurement of \ndi, \nni, and \noi\ by curve of growth and profile fitting techniques, as well as the analysis of interstellar molecular hydrogen.  This \h2\ model is used in the \di, \ni, and \oi\ analysis.  The paper concludes in \S 6 with  a discussion of the results and a comparison with other recent D/H studies in the Galaxy.

\section{OBSERVATIONS}\label{obs}

\bd28\ was observed by \fuse\ through the LWRS aperture (30\arcsec$\times$30\arcsec) on 2000 June 13  for 2192 s (ObsID=M1080901), 2000 July 16 for 16677 s (ObsID=M1040101), and 2000 September 19 for 7864 s (ObsID=M1040105). One MDRS aperture (4\arcsec$\times$20\arcsec)  observation was made on 2000 July 17 for 24795 s (ObsID=M1040102).  The first observation was a short exposure to establish the feasibility and safety of using \bd28\ for the \fuse\ calibration program because its far-UV flux is very close to the bright limit of the \fuse\ detectors ($F_{\lambda} \sim 10^{-10} \ergcmsA$).  The subsequent observations were made for the purpose of testing the acquisition of data with the Focal Plane Assembly (FPA) offset to different positions along the dispersion direction as a technique to reduce fixed-pattern noise when the observations are later coaligned and added.  This procedure and the results are discussed below.  The \fuse\ instrument and its performance are described by Moos et al. (2000) and Sahnow et al. (2000).

The data were obtained in ``histogram'' mode because of the high far-UV flux of the star. In this data acquisition mode a two-dimensional
spectral image is accumulated on board, eliminating any timing information within individual exposures.  
The observations were split into four intervals
of approximately equal exposure time, each with a different X-offset of
the FPA.  The exposure time required for the last two LWRS
FPA positions was obtained in part during observation M1040101 and in part during
observation M1040105.  A shift in FPA X  causes
a corresponding shift in the location of the spectrum on the detector
along the dispersion direction.  The spectral shifts and  exposure times are given in Table \ref{fpsplit}. The FPA is fixed for the duration of each exposure so that there is no loss in spectral resolution relative to an observation taken at a single FPA location. The offsets for the MDRS observations span a smaller range in FES X than those of the LWRS observations in order to provide sufficient margin in the FPA range of motion. This margin is needed so that MDRS peakups can properly adjust the FPA's to correct for image motion caused by small rotations of the primary mirrors on orbital timescales. For further information on this image motion see Sahnow et al. (2000). The shifts caused by this image motion, together with shifts caused
by similar rotations of the gratings, result in a distribution of shifts 
clustered about the mean offsets given in Table \ref{fpsplit}.
Even for MDRS, these spectral offsets span a range roughly twice 
the width of typical interstellar absorption lines or of
the high-frequency components of the detector fixed pattern noise.

Exposure durations varied from 430 s to 535 s in order to minimize Doppler smearing of the spectrum by satellite orbital motion so that there is no degradation of spectral resolution. Most of the LWRS exposures were obtained during orbital night when \oi\ and \ni\ airglow emission is absent and that of \hi\ is significantly reduced.  When exposures having both orbital night and day segments are considered, over 70\% of the LWRS data was obtained during orbital night.  Over 82\% of the MDRS data was obtained at night.

The data were processed with the latest version of the FUSE calibration pipeline ({\tt CALFUSE} v. 1.8.7).  For each channel, data for each exposure were aligned by cross correlation and combined on a common heliocentric wavelength scale.  Part of the resulting spectrum for the LWRS observations is shown in Figure \ref{bdspec}.  The LWRS LiF co-added spectra have S/N$\sim 100$ per 0.05 \AA\ resolution element in the LiF channels and S/N$\sim 60$ in the short-wavelength silicon-carbide (SiC) channels.  The S/N in the MDRS spectra of \bd28\ are about 25\% lower than for LWRS. These S/N levels are close to the photon-noise limits.  The analyses presented in this paper uses primarily SiC channel spectra.   Figure \ref{lyspec} shows small sections of the spectrum  near several of the Lyman lines and demonstrates the high quality of our spectra.  Our analysis of  spectral features in the \bd28\ co-added spectra indicates that the spectral resolution is consistent with $\lambda/\Delta\lambda \sim$ 20,000, the nominal spectral resolution obtained by the \fuse\ instrument (Sahnow et al. 2000).  It is evident from Figures \ref{bdspec} and \ref{lyspec} that airglow has a negligible effect on these spectra.  \lyb\ is the strongest airglow transition in the \fuse\ band pass and the observed \lyb\ profile is only  weakly affected.  \oi\ airglow, if present, would primarily effect the stronger transitions longward of $\sim950$ \AA, but the \oi\ column density is determined from much weaker lines that are unaffected.

Figures \ref{bdspec}-\ref{model923} show that the saturated \hi\ lines have non-zero residual flux (on the order of a few percent).   The \fuse\ instrument has very low levels of scattered light (Moos et al. 2000, Sahnow et al. 2000), as shown by the zero flux level below the Lyman limit (see Fig. \ref{bdspec}). The residual flux in the Lyman lines is probably the result of wings of the instrumental line spread function (LSF; see Kruk et al. 2001).  The shape of the LSF and its potential effects on the line profile analysis is discussed further in \S\ref{dnointro}.

\section{STELLAR MODEL FOR BD +28$^{\circ}$ 4211}\label{bd-model}

MacRae, Fleischer, \& Weston (1951) reported the first spectral
analysis of \bd28\ and classified it as an extremely blue star with
peculiar spectrum (Op).  The optical spectrum of \bd28\ attracted
their attention because of its very blue continuum and the presence of
Balmer and \ion{He}{2} lines.  Greenstein (1952) estimated the
distance to \bd28\ and found out that the star was a subdwarf.  \bd28\
is now classified as sdO, based on the study of Moehler et al.\
(1990), in which the class is defined as having \heii\ and
strong Balmer absorption lines.  Although low resolution optical
spectra show only these lines, high-resolution spectra reveal, in
addition, many weak metal lines.  Herbig (1999) obtained
high-resolution optical spectra of \bd28\ at the Keck~I telescope and
detected a significant number of narrow absorption and emission lines
(\ion{C}{4}, \ion{N}{4}, \ion{N}{5}, \ion{O}{4}, \ion{O}{5}, and
\ion{Si}{4}), and also emission cores in H$\alpha$, H$\beta$, and
\ion{He}{2} $\lambda\lambda$6560, 5411, 4685 lines.  Using the
\ion{O}{4} absorption lines, Herbig (1999) set an upper limit on the
star's rotational velocity, $v \sin i \leq 4\ \rm{km}\,\rm{s}^{-1}$.
Even though \bd28\ is a well-known spectrophotometric standard star in
the ultraviolet and optical (see, e.g., Bohlin 1986; Massey et al.
1988), its atmospheric parameters were investigated only recently by
Napiwotzki (1993), Werner (1996), and Haas et al.  (1996).

Using H+He NLTE stellar model atmospheres, Napiwotzki (1993) tried to
fit the Balmer line profiles in \bd28's optical spectrum with no
success.  He demonstrated that the effective temperatures obtained
from fits of individual Balmer lines showed large discrepancies.  This
inconsistency, known as the Balmer line problem, was also observed in
central stars of old planetary nebulae and DAO white dwarfs (see,
e.g., Napiwotzki \& Sch\"onberner 1993; Napiwozki \& Raunch 1994;
Bergeron et al.\ 1994).  To overcome this problem, Napiwotzki (1993)
modeled the \ion{He}{1} $\lambda$5876 line, as it is a very sensitive
diagnostic of effective temperature (\teff), and obtained $T_{\rm{eff}} = 82$,000~K
and $\log({\rm{He/H}}) = -1.0$.  He also derived the gravity $\log g =
6.2$ from the higher Balmer lines.

By analyzing the optical spectra of \bd28 and the central star of the
planetary nebula S216, Werner (1996) concluded that the Balmer line
problem was caused by the omission or inadequate inclusion of metal
opacities.  For instance, he matched the Balmer lines
(H$\alpha$-H$\delta$) of \bd28 combining Napiwotzki's (1993)
parameters (\teff $ = 82$,000~K, $\log g = 6.2$, and $\log
{\rm{(He/H)}} = -1.0$) and the improved Stark-broadened CNO line
profiles.  He demonstrated that the effect of this added opacities
increased the temperature in the deeper layers of the atmosphere, and
decreased the temperature in the superficial layers, therefore
modifying the Balmer line formation.  He noted that the low Balmer
series members cores are deeper as they are formed higher in the
atmosphere, where the temperature is lower.  On the other hand, the
high Balmer series members were less affected because they originate
from deeper layers of the atmosphere.  This new temperature structure
caused by the addition of more realistic metal line opacities resolved
the Balmer line discrepancy.  Interestingly, Bergeron et al. (1993) had
reached a similar conclusion to that of Werner (1996) by analyzing the
DAO white dwarf Feige~55 and using a LTE iron-blanketed model.

For the purpose of this study, we adopt the stellar atmospheric
parameters determined by Napiwotzki (1993) as they are the best
available.  We compute a grid of H+He NLTE model atmospheres with the
following atmospheric parameters:  $T_{\rm{eff}} = 78$000~K, 82000~K,
and 86000~K; $\log g = 6.0$, 6.2, and 6.4; and $\log ({\rm{He/H}}) =
-0.6$, $-1.0$, and $-1.4$.  We also compute a metal-line blanketed
model including the following chemical composition:  H, He, C, N, O,
Si, S, Fe, and Ni.  We then compute a grid of synthetic spectra
from \lya\ to the Lyman limit using the
TLUSTY program developed by Hubeny \& Lanz (1995) and an upgraded
version of the program SYNSPEC (I.~Hubeny 2000, private
communication), which incorporates the Lemke (1997) Stark broadening
tables for hydrogen.  We explored the model parameter space 
(\teff, $\log g$, and $\log({\rm{He/H}})$) to evaluate the
uncertainties in the stellar \hi\ Lyman + \heii\ Balmer line profiles.  
\S4 describes in
more details the use of stellar models to dermine the \hi\ column
density.

The \fuse\ spectrum of \bd28\ (Fig. \ref{bdspec}) shows many stellar and interstellar
lines.  The stellar spectrum displays the Lyman series of hydrogen
from \lyb\ up to Ly-9 and the \heii\ Balmer series from 1084 \AA\ up
to at least 942 \AA.  The spectrum also contains many stellar lines of
highly ionized species, such as \ion{C}{4}, \ion{N}{5}, \ion{O}{4},
\ion{O}{5}, \ion{O}{6}, \ion{Si}{4}, \ion{P}{5}, \ion{S}{4},
\ion{S}{5}, \ion{S}{6}, \ion{Fe}{6}, \ion{Fe}{7}, \ion{Ni}{6}, and
\ion{Ni}{7}.  The strongest lines are the \ion{N}{4} $\lambda$923.15
sextuplet, \ion{S}{6} $\lambda\lambda$933.38, 944.52 doublet,
\ion{N}{4} $\lambda$955.33, and \ion{O}{6} $\lambda\lambda$ 1031.91,
1037.61 doublet.  In order to identify potential blending of
photospheric and ISM lines, we first performed an abundance analysis of
\bd28's atmosphere and then calculated a synthetic spectrum
using the derived abundances.  Figure \ref{model923} illustrates
the comparison between the synthetic spectrum and the FUSE spectrum
for the wavelength range 920--927 \AA.

\bd28\ has several strong, narrow metal lines in the \fuse\ range and that
made it possible to register the wavelength scale of the stellar model with the observed spectrum.  In particular, \svi\, \wl944.523 is
seen in both SiC channels and has a deep, narrow core.  We used this line to align the model with the spectra. As expected, the zero
point wavelengths are different from channel to channel because it depends on the FPA position as well as the location
of the star in the aperture. It varied between $-60 \kms$ and $+60 \kms$ in
our data, although larger offsets were occasionally detected in the first year 
of \fuse\ operations (Sahnow et al. 2000). Once the wavelength offsets are corrected, the
best estimate from the \fuse\ data
of the  velocity of the star relative to the ISM is $+34 \pm 6 \kms$, a value that is consistent with our more accurate determination of $+31\pm2 \kms$ from \hst\ spectra (see below). We increased the error to $\pm8 \kms$ to take into account any
remaining uncertainties in the \fuse\ wavelength calibration.

\section{DETERMINATION OF THE \hi COLUMN DENSITY}\label{nhi}

\subsection{Hubble Space Telescope Observations}

The determination of the interstellar \hi\ column density along the sight line to \bd28\ uses an extensive series of archival observations taken 
with the Space Telescope Imaging Spectrograph (STIS) on board the
\hst.  The \hi\ column density may in
principle be determined using the \fuse\ observations of this star.
However, most of the Lyman-series lines in the \fuse\ bandpass are on
the flat part of the curve of growth. In addition, the damping wings of interstellar \lyb\ and \lyg, the strongest \hi\ transitions in the \fuse\ bandpass are significantly less prominent than those for \lya\ and comparable in strength with the stellar \hi\ + \heii\ profiles.  Therefore,
the only reliable  procedure for determining an accurate \hi\ column density to
\bd28\ is by fitting the Lorentzian wings of the very strong 
\lya\ line.  As illustrated below, the interstellar \hi\ profile is much stronger that its stellar counterpart.

\bd28\ is a calibration source for STIS, and observations have
been acquired over several years with the E140M echelle grating
projecting the two-dimensional spectrum onto the FUV MAMA (Kimble et
al. 1998; Woodgate et al. 1998).  All data were taken through the
$0\farcs2\times0\farcs2$ aperture and cover the spectral range
$\sim1150 - 1750$ \AA.  This instrumental setup provides a spectral resolution
$R\equiv \lambda/\Delta \lambda \sim45,000$ ($\Delta v \sim 7 \kms$ FWHM), although the line spread function for data taken through
this aperture has significant power in broad wings that extend well
beyond this width (see Figure 13.87 in the Cycle 9 STIS Instrument
Handbook).  We used  eight such observations taken over the time
period 1997 Sept. 19 to 2000 Nov. 10.  Exposure times for the
individual observations were in the range 350--1800 s.  Figure
\ref{lyalpha} shows a coaddition of the data sets used in our
analysis in the region near \lya.  This figure demonstrates the
quality of the data and shows the importance of high spectral
resolution for separating the numerous stellar and interstellar lines in this region.

\subsection{Data Calibration and Reduction}

The E140M data have been reduced using the CALSTIS pipeline by two different approaches in an attempt to uncover sources of systematic error. One used the pipeline as developed by the STIS
Instrument Definition Team, the other the standard STSDAS
pipeline distributed within IRAF (see Voit 1997).  The CALSTIS
pipeline removes an estimated dark level, applies a flat field,
corrects small-scale distortions, identifies the spectral trace,
removes scattered light, extracts the one-dimensional spectrum, and
performs the wavelength and flux calibrations. We show below that both analysis approaches yield the same result.

The scattered-light removal process is among the most important,
particularly for our purposes where we are interested in the
large-scale wings of the \lya\ profile.  Because the strength of the
echelle-scattered light can be significant in
E140M spectra, the Bowers \& Lindler algorithm
(see Landsman \& Bowers 1997; Bowers et al. 1998) was used to estimate and
remove scattered light from the data in the pipeline procedures.
We  found that this algorithm does a good job of determining an
appropriate zero-level, as evidenced by the small residual fluxes in
the saturated cores of strong lines (e.g., the \lya\ profile shown
in Fig. \ref{lyalpha}).

After extracting the spectrum, several echelle
spectral orders in the regions adjacent to \lya\ were combined using a weighted
averaging scheme where the orders overlap.  The flux calibration
between orders seems good for the E140M data sets used here.  Even so,
there can be some small fluctuations in the final co-added spectrum
due to mismatches at the edges of orders, although in general these can
be masked out and do not significantly affect our analysis.

\subsection{Analysis Methodology}

The method for deriving the \hi\ column density along this sight line
follows Jenkins (1971) and subsequent work (see discussion,
particularly of error analysis, in Vidal-Madjar et al. 1998; Howk,
Savage, \& Fabian 1999; Jenkins et al. 1999; Sonneborn et al. 2000).
The distribution of optical depth, $\tau(\lambda)$, with wavelength
for \lya\ is given by the product of the absorption cross-section,
$\sigma(\lambda)$, and the column density.  For \hi\ at \lya\ this is:
\begin{equation} 
\tau(\lambda) \equiv \sigma(\lambda) N = 
  4.26\times 10^{-20} N (\lambda - \lambda_0)^{-2},
\end{equation}
where $\lambda_0$ is the \lya\ line center at the velocity centroid of
interstellar hydrogen.  The \hi\ column density is determined to be
the one that best matches the above distribution of optical depth with
the observed profile.  This method only gives information on the total column
of \hi, since the details of the component structure are all within
the strongly-saturated core of the \lya\ line which spans a velocity range of $\pm200 \kms$. It is extremely unlikely that the damping wings are affected by low column density \hi\ absorption features. Any \hi\ absorber that could modify the shape of the damping wings would have to have a heliocentric velocity $|v|> 400 \kms$ (see Fig. 9 of Sonneborn et al. 2000).  The broad damping wings of \lya\ are not sensitive to the Doppler broadening parameter.  We therefore have very little information about $b$ from the \lya\ analysis.

The derivation of \nhi\ was performed with two independent,
though similar, fitting procedures.  Both vary the important free
parameters, which for our purposes include $\lambda_0$ and
\nhi, to minimize the $\chi^2$ statistic between the model,
after convolution with the instrumental LSF, and the
observed spectra (see, e.g., Jenkins et al. 1999).  We note that the
specifics of the LSF are relatively
unimportant for \nhi\ as large as is found on this sight line (see, e.g., Appendix A of Howk et al. 1999).  This insensitivity includes the possible
presence of the broad wings of the STIS E140M LSF mentioned above, which
are on a much smaller scale than that of the \hi\ absorption.

We applied masks to exclude obvious stellar and interstellar absorption features from the
fit.  We also avoided fitting the very deepest portion of the \lya\ line core
(effectively those regions with flux $<10\%$ of the stellar
continuum within $\sim\pm2$ \AA\ of the line center) because of the
possibility that stellar lines may be unseen in this low
signal-to-noise region of the spectrum.

The stellar atmosphere for this type of star  also has
broad \lya\ + \heii\ absorption.  It is, therefore, important to account for
the stellar continuum underlying the interstellar absorption in our
fitting process.  For this purpose we used the synthetic spectrum and stellar \lya\ + \heii\ line profile described in \S \ref{bd-model}.    

The computed stellar spectrum was shifted to a heliocentric velocity of
$+20.2 \kms$, the radial velocity of \bd28, for comparison with
the observed \lya\ profile.  We  determined this radial velocity by
measuring the velocities of prominent stellar lines in the STIS E140M
observations, the average of which gives a heliocentric velocity of
$v_{hel} = +20.2\pm0.8 \kms$.  The quoted uncertainty is the
$1\sigma$ standard deviation of the individual measurements about the
mean and does not include the uncertainties in the STIS absolute
wavelength scale, which probably correspond to $\sim1.6$ to $3.2 \kms$.  This determination agrees well with the stellar radial velocity of $v_{hel} = +21.4\pm1.7 \kms$
derived by Herbig (1999), where we use the mean
and $1\sigma$ standard deviation of his individual measurements.  For comparison, the average velocity of
interstellar material along this sight line is $v_{hel} \sim -9$ to
$-13 \kms$ (stronger lines give more negative velocity components due to the presence of weaker components on the blue side of the interstellar profiles).  The stellar radial velocity is well constrained, although we found that offsets from this best value as large as $5 \kms$ have no effect on the derived \hi\ column density.

The model stellar spectrum was scaled to best match the observed flux distribution using a second-order Legendre polynomial, the parameters
of which are treated as free parameters in the fitting process.  In
general, the shape of the calculated stellar model far from line center fits the data relatively well.  
However, the adoption of a low-order polynomial correction to the
synthetic stellar  continuum  allows for several
possible systematic effects, including  mis-matches between the observed
spectrum and the model, slight redistribution of flux in
the scattered light removal process, stellar atmospheric variability,
and other properties of the detector and observations that could
change the absolute flux distribution on Angstrom scales.

We  also allowed for the effects of uncertainty in the flux zero point,
treating it as a free parameter in the fit.  This corrects for
any small uncertainty in the background subtraction on scales of
several Angstroms.  The derived corrections to the flux zero point
were always less than a percent or two of the continuum away from the line core.

Because of the possibility that the stellar  \lya\ profile could be  variable, we  fit separately each of the
individual observations taken at different times. No evidence of variability was found.  This is
discussed in more detail below in our discussion of systematic
uncertainties.

\subsection{Results and Discussion of Systematic Uncertainties}

Figure \ref{lyalpha} shows the best fit interstellar \hi\ profile on top of the
coadded STIS observations.  The final estimate for the \hi\ column is
$\log N(\mbox{\hi}) = 19.846^{+0.036}_{-0.034}$, where the
uncertainties are $95\%$ confidence limits ($2\sigma$ if Gaussian
errors are appropriate) including systematic effects.  This is an average of the results from the two \nhi\ determinations,
which differed by only 0.003 dex ($<1\%$), and uncertainties, for
which we  adopted the larger of the uncertainty estimates.

The uncertainties in \nhi\ given above include  estimates of the
statistical (random) and systematic uncertainties added in quadrature.
Because of the high quality of the data and the strength of the \lya\ 
damping wings, the systematics dominate the statistical uncertainties in our analysis.  Given
the importance of the systematic uncertainties, we discuss several
below and their impact on the $N(\mbox{\hi})$ determination.

{\em Stellar model uncertainties --} The largest identified source of
systematic uncertainty for \nhi\ is the adopted stellar model.
However, fits to the observed \lya\ profile which use only a straight
line continuum rather than a stellar model for normalization produce
\hi\ column densities that are larger by only +0.1 dex ($\sim25\%$).  This 
sets an upper bound on the magnitude of the uncertainties due to the
stellar models.

The effects of imperfectly-known stellar atmospheric
parameters on \nhi\ were estimated by refitting the interstellar \lya\
absorption profile using model atmospheres with the most extreme
stellar properties that are still nominally consistent with the
determination of the atmospheric parameters (see \S \ref{bd-model}).  For this
purpose we chose those models with the strongest and weakest stellar \lya\ 
absorption profiles from a grid of atmospheres covering a wide range in
parameters: $\log {\rm He/H} (-0.6$ to $-1.4$), $T_{eff}$ (78,000 to
86,000 K), and $\log g$ (6.0 to 6.4).

The \nhi\ values derived in this way were treated as
extrema corresponding to 95\% confidence limits ($2\sigma$ assuming
Gaussian statistics) about the best-fit result.  While this approach
does not account for any uncertainties in our knowledge of the
fundamental physics of stellar atmospheres or radiative transfer, the adopted extreme
atmospheres encompass a sufficiently large range of parameter space
that the effects of any subtle atmospheric physics are expected to be
small in comparison to the uncertainties estimated in our approach.
The uncertainties associated with the adopted stellar model  and our
allowance for stronger and weaker stellar \lya\ profiles dominate the
uncertainties in the \nhi\ determination.

We note that the effects of line-blanketing within the model
atmosphere will be much smaller than the effects of the extreme
atmospheric parameters investigated above.  We  verified this assumption with
appropriate model calculations, and that the
variation in the stellar \lya\ profile  occurs primarily in the core of the line where the
residual flux in the observed spectrum is zero.

{\em Residual scattered light --} In principle, residual scattered
light in the STIS E140M spectra could bias the \nhi\
determination.  The core of \lya\  is indeed very near zero,
and we  allowed the true flux zero point to vary as a free parameter in
the fits.  However, if light were distributed in the line wings in an
appropriate way during the scattered light removal process, this
could potentially be a source of systematic uncertainty in our \hi\
column density determination.  

As a check of the \nhi\ value derived from  STIS E140M observations,
we  also analyzed archival observations taken with the pre-COSTAR
Goddard High Resolution Spectrograph (GHRS) first-order G160M grating
through the small science aperture ($0\farcs25 \times 0\farcs25$).
These spectra have a resolution of $\sim15,500$ ($\sim19.3$ km s$^{-1}$ FWHM).  While the observations do not cover the entire extent of the red wing
of \lya, the holographically-ruled G160M grating had excellent
scattered light properties (Cardelli, Ebbetts, \& Savage 1993).  
Eight observations with a total exposure time of 1152 s were reduced and combined using procedures described by Howk et al. (1999).   From these spectra we found  $\log N(\mbox{\hi}) = 19.842\pm0.020$ ($2\sigma$,
statistical uncertainties only).  This value is so close to the value
derived using the STIS E140M observations that we believe residual
scattered light uncertainties are not likely to be a significant
source of error.

{\em Stellar atmospheric variation --} The atmospheric profile of the
star could be variable, and there is some evidence that
\bd28\ may be part of a binary system (see Napiwotzki 1999; Massey 
\& Gronwall 1990).  To account for the possibility that the stellar
\lya\ profile might vary, we separately  calculated the best-fit \hi\ 
column densities for each of the eight individual STIS observations.  The dispersion in these best-fit measurements for the
ensemble of observations was small and was added in quadrature to the
final uncertainty estimate.  The dispersion of the individual
measurements about the mean was much smaller than that resulting from the variation of stellar model parameters described above.

\section{COLUMN DENSITIES OF ATOMIC DEUTERIUM, NITROGEN AND OXYGEN}\label{N(DI)}
\subsection{General Considerations}\label{dnointro}

Both curve of growth (COG) and  profile fitting techniques were used to determine \ndi\ and \noi\ in order to better understand potential sources of systematic error. Profile fitting alone was used for \nni\ because there are only three unsaturated and unblended  \ni\ lines available (the \wl952 multiplet). The \fuse\ spectra were also analyzed to develop a model for interstellar \nh2\ for rotational levels $J=0-4$.  The comparison of measurements from multiple channels (SiC1 and SiC2) and observations (LWRS and MDRS) is a powerful tool that  improves our statistics and  helps to identify poor data quality due to instrumental effects. 

In the COG method the equivalent width, \eqw, of several lines of the
same species having  different oscillator strengths are measured
and their distribution in the $\log(W_{\lambda}/\lambda)-\log(Nf\lambda)$ plane is compared with a
theoretical curve of growth for a single absorption component subject to a Maxwellian
velocity distribution.  The free parameters in the fit are the column density, $N$ and Doppler spread parameter, $b$. However, if there is a complex velocity
distribution, $b$ is regarded
as an ``effective'' Doppler spread and as such has no physical meaning.
Continuum placement and equivalent width measurements are treated 
in independent steps and assume some a posteriori conservative error bars 
to take into account the uncertainties in continuum placement. The COG
method is potentially vulnerable to error because of the assumption of a single-component profile. In particular, at the \fuse\ spectral resolution narrow, unresolved components could be
concealed by broader features, which could result in underestimated column densities (Nachman \& Hobbs 1973, Savage \& Sembach 1991).  Jenkins (1986) showed
that column densities accurate to $\sim$10\% can be obtained with the
COG method provided that the lines are not heavily saturated
$(\tau_0 \le 2)$ and have smooth distributions of $b$ and $\tau$ in their components.
In the case where all transitions are on the linear part of the
COG, the column density is insensitive to the line-of-sight velocity structure.  One of the advantages 
of the COG method is that it is not too sensitive to the LSF. On the other hand, transitions blended with lines from other species should
 be avoided in this analysis technique. In this case, profile fitting may be more appropriate.

The profile fitting method makes the same basic assumptions as
the COG: an interstellar absorption line is formed in a single component and is
described by two parameters: $N$ and $b$. In addition the wavelength centroid of the line, or radial velocity, $v_r$, is also a free parameter.  The fitting technique also
assumes  that each interstellar component produces absorption
lines described by Voigt profiles.  We  note that in our analysis the Doppler parameter, $b$, has no unique physical interpretation.  Normally $b^2 = 2kT/m + v_{turb}^2$.  However, from the \fuse\ spectra alone, with spectral resolution of $\sim 15 \kms$, temperature ($T$) and turbulent motion ($v_{turb}$) effects on the line widths are not easily separated.  We can, however, use $b$ to set an upper limit on $T$.  

Lines of sight with multiple components are often more reliably analyzed with a fitting technique
than it is with the COG. In practice, to fit a multiple-component COG
requires a prior knowledge of $v_r$ and $b$ for
each component as well as their expected relative ratios. This
information is not necessary with profile fitting because in principle it lies in the profiles.  For the same reason, the blending of adjacent lines is also
easier to model with a fitting procedure. 

A second significant advantage of our profile fitting analysis is
that absorption lines and continua are
fit simultaneously so that continuum shape and placement are treated as 
free parameters. We have used the profile fitting program {\tt Owens.f} (Lemoine et al. 2001, H\'ebrard et al. 2001). {\tt Owens.f} is a powerful profile fitting program that simultaneously models multiple spectral lines, species, and absorption components in an arbitrary number of spectral regions, or windows.  Many quantities characterizing the fit may be treated as free or fixed parameters ($N, T, v_{turb}$, LSF, continuum placement and shape, background level, velocity offset between spectral windows, etc.). The program uses an optimized \chisq-minimization algorithm to obtain fast convergence.  The flexibility of {\tt Owens.f} allowed us to analyze many different cases.  As with any convergence algorithm, parameter space was explored to verify that a converged solution was the true minimum and not a local minimum far from the true solution.
 This was done by changing
the initial parameters and/or conditions and then running the program again to reach a new converged solution. A potential disadvantage of profile fitting is that knowledge of the
instrument LSF is required, and errors in the assumed LSF have the potential
to bias the results.

The instrumental LSF is a key parameter in the profile fitting analysis of interstellar absorption lines.  The characterization of the \fuse\ LSF was incomplete at the time of this analysis (see Kruk et al. 2001).  Although there is evidence that the \fuse\ LSF has wings that effect the zero level of heavily saturated lines (\S\ref{obs}), two independent analyses (H\'ebrard et al. 2001, Wood et al. 2001) show that modelling of weak unsaturated lines with a single Gaussian LSF and a double Gaussian LSF give very consistent results within the uncertainties of the \fuse\ LSF parameters (see the companion papers referenced above for further details).  For our profile fitting analyses we used a single gaussian LSF for each component modelled.

In the next subsection the interstellar \h2\ model is derived. The following subsections describe the COG analysis of \di\ and \oi\ and the profile fitting analysis of \di, \oi, and \ni.

\subsection{Interstellar \h2\ Column Densities}\label{h2model}

The \fuse\ spectra of \bd28\ contain a large number of absorption lines from interstellar molecular
hydrogen. Absorption arising from \h2\ rotational levels $J=0-4$ can be clearly
seen along this line of sight.  Many \h2\ lines are blended with
stellar and other ISM features. For transitions corresponding to the rotational level $J=4$
there are not enough unblended lines to derive a reliable column density using either COG or
profile fitting techniques.  However, a $3\sigma$ upper limit for \nh2\ $J=4$
is derived. A few \h2\ transitions from $J=5$ and 6 may also be
present. The best candidate is Werner $0-0$ Q(5) at 1017.831 \AA. However, the \fuse\ spectra of \bd28\ contain many unidentified stellar
features and it is possible that this and other weak transitions could be stellar in
origin, or if actual \h2\ ISM lines, blended with stellar lines. Due to
these uncertainties, the column densities for $J \ge 5$ could not be estimated. The \h2\ column densities for  rotational levels $J=0-3$ were
measured using both profile fitting and COG methods, using 
lines that appeared free of blending, as judged by the shape of the absorption
profile and with the help of the stellar model.

Profile fitting used only unsaturated lines, i.e. lines for which the absorption profile
deconvolved with the LSF did not have
a residual intensity in the line core $< 0.1$. The column density for each rotational level was determined independently.  The synthetic stellar spectrum gave guidance when selecting the \h2\ lines to be analyzed in order to avoid blending with stellar features.  Each spectral region (several
Angstroms) was first normalized with a spline function.  A subset of this  region, a spectral window of $\sim1$\AA, was used by {\tt Owens.f} 
 where a polynomial (up to
$4^{th}$ order) was fit to the continuum away from the line.  We restricted the size of the spectral windows used in the {\tt Owens.f} analysis because of concern about small nonlinearities in the \fuse\ wavelength calibration on larger scales. The number of lines analyzed for each $J$ level are listed in Table \ref{nh2table}. All the \h2\ lines analyzed
lie away from the \hi\ lines. The background was set to zero for
each window. The width of the Gaussian LSF was free to vary from window to window and between
successive fits.  Once the best fit was found,
corresponding to a minimum \chisq, the
error in the column density was calculated by fixing \nh2\ to a different value and computing a new optimized fit, which always had a larger \chisq.  The difference of the new reduced \chisq\ from \chisqmin\ determined the 
  1, 2, 3, 4, 5, 6, and
$7\sigma$ deviations from the best fit.  The adopted $1\sigma$ error was the average of these values to allow for any asymmetry in the $\Delta$\chisq\ curve. Only an upper limit was derived for  \nh2\ $J=4$, the result of its weak transitions being indistinguishable from weak stellar features. \nh2\ and $2\sigma$ errors derived this
way are given in Table~\ref{nh2table}.

The \h2\ equivalent widths were measured by directly integrating the area of several lines for each rotational level for $J=0-3$.  The continuum near each line was defined by a Legendre
polynomial up to $5^{th}$ order.  \eqw\ of each transition is the weighted average of \eqw\ measured in each channel where
the line was present. 
A  single-component Gaussian COG was fit to the data for each rotational level, and the residuals about the best-fit curve were
minimized (Figure \ref{h2cog}).   The resulting column densities are completely consistent with the profile fitting results, although with somewhat larger uncertainties.  Assuming the population of the
lower rotational states of \h2\ is determined by collisional
excitation and a Boltzmann distribution, the excitation temperatures are $T_{1,0} = 313$ K , $T_{2,1} = 294$ K
 and  $T_{3,2} = 296$ K. However, the similarity of $T_{i,j}$ for $J=0-3$ indicates that one or more of these assumptions may not be correct. The uncertainty in \nh2\ $J=0$ may permit a more typical  value of $T_{1,0} \sim 100$ K.  The total \h2\ column density is log(\nh2)$=15.13^{+0.20}_{-0.08} \cmsq$.

\subsection{\di\ and \oi\ Curve-of-Growth Analyses}\label{dicog}

The high S/N of the \fuse\ observations of \bd28\ permits analysis of
the \di\ transitions down to 916 \AA\ where the \hi\ profiles
start to overlap.  However, some of the higher \di\ Lyman series lines (916--918 \AA)  have strengths that vary significantly from line to line and channel to channel. These inconsistencies probably
originate in the fixed pattern noise of the detector and/or in blends with weak stellar features or interstellar \h2. Therefore, the COG analysis of \di\ was limited to the six transitions listed in Table \ref{doeqw}.   Most of the \di\ lines (\lyg\ -- \lyeta) lie in the blue wing of the corresponding \hi\ lines. 
Four out of the six
transitions  fall on the linear part of the curve of growth. Hence, the
velocity structure on the line of sight does not influence the final column density estimate (see \S \ref{diprof}).

\subsubsection{\di\ Continuum Normalization}

The stellar \hi\ and \heii\ and interstellar \hi\ absorption profiles dominate the shape of the
continuum near the \di\ feature.  A continuum model was constructed, where the key parameters are \nhi, \nh2, and $b$ for the ISM model computed with {\tt Owens.f} and
\teff, $\log g$ and He/H for the stellar model. We  adjusted the stellar plus interstellar model to fit the
data (a small zero-point shift in the model's wavelength scale and scaling the absolute flux of the model to match the observed flux level -- see Fig. \ref{model923}). These are discussed further below. 

The stellar model has been discussed in detail in \S \ref{bd-model}.  The 
uncertainties in \teff\ and $\log g$ considered in the \nhi\ analysis ($\sim$5\%) have a negligible effect on the shape of the \hi\ + \heii\ 
profile near the \di\ lines we are studying. While varying these parameters would change the goodness
of the fit far from the core of the \di\ lines, the continuum
placement  would be similar.  As to the He/H ratio, this  affects only the two
strongest lines (\di\ \lyg\ and \lyd) where \heii\ is readily apparent in the model. 
 
The overall shape of the model agrees well with the
spectrum, although we noted some localized discrepancies. The different spectra showed flux scale differences from the model of up
to +5\% for LWRS and +10\% for MDRS. The latter suffered from small variations in slit transmission due to mirror motion (see Sahnow et al. 2000) and a small part of the flux was lost in this
process. The MDRS flux calibration is also  less well known
than that of the LWRS channel. 

We modelled the interstellar \hi\ profiles by using {\tt Owens.f} to fit  the relevant \hi\ Lyman series transitions (919 -- 972 \AA), forcing the column density to log(\nhi) = 19.85 and the adopted the \nh2\ parameters from Table \ref{nh2table}. A one-component fit agreed well with the data. A three-component fit was also performed (cf. \S\ref{diprof}) but there was no improvement in the fit because in \hi\ these components are buried in the saturated line cores.  The purpose of this \hi\ model was to find a good empirical match with the data  to assist in the measurements of \di\ \eqw, not to determine \nhi.  

\subsubsection{\eqw\ Measurements}

The general method for measuring \eqw\ for the \di\ lines is illustrated in Figure \ref{cogmethod}. 
First, the data were divided by the stellar model appropriately registered in \wl\ and $F_{\lambda}$ (Fig. \ref{cogmethod}a and d). This process sometimes produced what was judged to be a poor fit in the red wing of interstellar \hi, especially for \di\,\wl925. We believe this is  due to differences in the number, location,and strength of weak stellar lines  between the blue
and red wings.  In spite of the concern about such line blanketing differences, \eqw\ for \di\, \wl925 is consistent
with the other \di\ transitions analyzed. 

Second, the normalized spectrum was then corrected for interstellar \hi\ (Fig. \ref{cogmethod}b and e) by using the \hi\ model profiles described above.
\eqw\ of each \di\ feature was measured by fitting   a Gaussian to the residual \di\ absorption line (Fig. \ref{cogmethod}c and f) using all of the profile except
the far red wing of where the near-zero flux near the line core of \hi\ produces large residuals.
Since the principal source of error in this analysis is the stellar continuum
placement, we adopted a  conservative approach, scaling the continua by
$\pm$3\% and remeasuring \eqw.  This produced 
noticeably poor fits that represent extreme cases.  We assumed that
these fits were $2\sigma$ from the best fit.  Therefore, the \ndi\
errors include the systematic error due to continuum placement.

Table \ref{doeqw} gives the \di\ \eqw\ values for each channel and aperture. There are four independent measurements of W for each DI transition (SiC1 and SiC2, LWRS and MDRS).  These measurements were consistent to within $1\sigma$, except for the three cases where \eqw\ was omitted.  These
measurements were discarded because they were inconsistent (discrepant by $\ge$50\%) with the other
measurements of the same transition.  We concluded that these cases were probably caused by detector artifacts 
(fixed-pattern noise) after careful comparison of line profiles in the four aperture and channel combinations. The far right hand column in Table \ref{doeqw} is the mean \eqw\ used in the COG.  The COG fit for \di\  is shown in Figure \ref{diallcog}. We found log \ndi$= 14.99 \pm 0.10 ~(2\sigma)$ and $b=5.2 \kms$.

\subsubsection{\oi\ Curve of Growth}

A curve of growth was constructed for \oi, for each segment (SiC1 and SiC2,
MDRS and LWRS), using the lines and $f$-values of Table \ref{doeqw}. \oi\ \wll 924.52, 936.63, and 950.88 were not used because of blending with stellar or \h2\ features. The general
procedure to construct the curve of growth is similar to what was described
above for the \h2\ COG analysis (\S \ref{h2model}). For the \oi\ COG we chose lines that are free of blending with other
ISM or stellar features.  Unlike \di\ and \h2,  \oi\ $f$-values may have relatively higher uncertainties in the \fuse\ bandpass,
all of them coming from theoretical calculations (Morton 1991; Morton 2000, private communication; Verner, Barthel, \& Tytler 1994).
Unlike the pathological case of \oi\,\wl1026.5 pointed out by Jenkins et al. (2000), the other \oi\ $f$-values
appear to be self-consistent. \eqw\ and associated errors
measured for the four spectra are listed in the Table \ref{doeqw}.  The COG fit to the mean \eqw\ is shown in Figure \ref{oiallcog} and the derived column densities are given in Table \ref{oitable}.  A COG fit to the weighted
average of \eqw\ for the different channels and apertures yields log \noi$_{COG} = 16.23 \pm 0.08 \cmsq~(2\sigma)$ and $b=4.5 \kms$.

\subsection{Profile Fitting Analysis}\label{diprof}

A profile fitting analysis of \di, \ni, and \oi\ column densities was performed to provide a comparison with the results of the COG analysis in the previous subsection. 
The general technique is identical to that discussed in \S \ref{h2model}.  Since {\tt Owens.f} was used to fit \di, \ni, and \oi\ simultaneously, the following discussion also applies to \noi\ and \nni.

The five \di\ transitions analyzed  (\lye, \lyz, \lyeta, \lyth, and \lyi)  are not saturated.  
With the use of a fitting
procedure, the choice of the lines to analyze is different than with the COG technique. An
example is \lyz\ which was not used in the COG analysis because
it is  blended with \h2\ $J=1$ and $J=2$. By adopting the \h2\ solution (Table \ref{nh2table}) as a constraint in the fit, this transition is now part of the fitting analysis and easily deblended from the \h2.  We note that the errors in \nh2\ are not significant for the \ndi\ analysis because the \h2\ lines blended with \lyz\ are on the flat part of the COG. The two strongest \di\ transitions used for the COG (\lyg\ and \lyd) were omitted from the profile fits because
they are mildly saturated ($\tau_0 \sim 2$).  Uncertainties in our knowledge of the \fuse\ LSF could influence the results derived from profile fits to
such lines. In any case, \ndi\ is well constrained by the other
five transitions.

Although the continuum close to the \oi\ lines is very smooth and easily fit with a low-order polynomial, some of the  \oi\ transitions used in the COG are indeed  saturated
and for this reason they were not included in the \noi\ profile fitting analysis using {\tt Owens.f}. Only three \oi\ transitions were used to determine \noi: 919.92 \AA, 925.45 \AA, and 930.26 \AA. The strongest of these has $\tau_0 \sim 1$.

There are many \ni\ transitions in the \fuse\ range and four multiplets are  detected : \wll952, \wll953, \wll964, and \wll1134. The 952 \AA\ triplet is free of any
blending with \h2\ or ISM atomic lines and seems clear of
stellar features. Since all the lines of this triplet are still on the linear
part of the curve of growth, deriving the column density is straightforward.
This is not the case for the \wll953, \wll964, or  \wll1134. These 
multiplets ets are all  saturated and much less sensitive to $N$ than they are to $b$. The 964 \AA\ triplet is the next strongest after \wll952, but these lines are
blended with several \h2\ lines, \pii\, \wl963.8, and with unidentified stellar
features.  With only one multiplet (\wll952) we did not perform a COG analysis of \ni.  The \ni\ $f$-values used were provided by D. Morton (2000, private communication).

The {\tt Owens.f} analysis of the \di, \ni, and \oi\ lines allow many quantities that characterize the fits to be treated as free parameters.  These include the wavelength and flux zero points of each profile, the polynomial coefficients for each spectral window, the LSF parameters (discussed below), the column densities $N$, and Doppler parameters $b$.  In total, there were about 40 spectral windows and about 1500 degrees of freedom to fit 11 transitions.  The \h2\ column densities (Table \ref{nh2table}) were adopted as a fixed constraint. We fit the five lines of \di, three of \oi, and three of \ni\ in the four data sets simultaneously, providing a global solution.

The continuum for the \di\ line profiles was determined by a different approach than was
used for the \di\ \eqw\ measurements. There is sufficient information immediately
adjacent to the \di\  features to establish the continuum with a polynomial without normalizing the spectrum or modelling the stellar and interstellar \hi. 
The shoulder of the blue wing of the interstellar \hi\  profile, excluding the \di\ feature, was fit  by a $4^{th}$ order
polynomial.   Examples of the continuum and \di\ profile fits are shown in Figure \ref{difits}. The effective $b$ is not tightly constrained in this analysis because the true shapes of the \di\ profiles are 
unresolved for these weak lines. 

The effect of simultaneously modelling the interstellar 
\hi\ and \di\ profiles and fitting a polynomial to the stellar contribution alone was also examined.  
The line-of-sight velocity structure  becomes critical when higher Lyman series of \di\ and \hi\ are fit together. In this case for \bd28\ the effective $b$ is dominated by \hi\ and is  larger than the \di\ effective $b$, much larger than the expected $\sqrt 2$ difference between H and D.  This is probably the result of very weak \hi\ components not visible in \di\ or any other species that become a factor in the width high Lyman series profiles of \hi. Overestimating $b$ for \di\  leads to an anomalously low \ndi.   When \di\ and \hi\ are fit together a larger fitting
window ($\Delta\lambda\sim$4 \AA) is also required, forcing \di\ to be shifted
from its true center as a result of small residual nonlinearities in the \fuse\ wavelength calibration.  Due to the introduction of these systematic errors we rejected this fitting approach in favor of the techniques described in the previous paragraph and illustrated in Figure \ref{difits}. 

The influence of the LSF assumptions and the velocity structure were carefully examined. Three cases were considered in the profile fitting analysis.   One way
of investigating the sensitivity of the fit to the assumed LSF was to let {\it Owens.f} adjust it as part of the \chisq\ minimization. For Case 1 we considered a fixed LSF
associated with each channel/aperture combination, ignoring the possible variations with  wavelength.  The choice
of the LSFs was made after several {\tt Owens.f} test runs with different parameters. In Case 2, we assumed a LSF modeled by a single Gaussian with FWHM that varied freely with wavelength and from channel to channel. Case 3 examined the effect of line-of-sight velocity structure as derived from STIS \feii\ profiles on the derived column densities.  The continuum was determined in all three cases by the technique illustrated in Figure \ref{difits}.  For all three cases, the LSF was modelled as a single Gaussian for reasons discussed at the end of \S\ref{dnointro}.

A free LSF (Case 2) clearly
improved the fit, as indicated  in Table \ref{dnotable}, compared with the fixed LSF (Case 1). Only a few spectral windows showed  LSF width variations as large as 20\% ($\le2$ pixels).  Such variations may be the result of fixed pattern noise that is only partly corrected by means of the FP-split procedure, or from weak, unresolved photospheric features.  For Case 2  the reduced \chisq\ was slightly lower than before but the \ndi\ and \noi\ were unchanged.  The fit of the \ni\ lines, fit simultaneously with \di\ and \oi, showed slight differences but are within the $2\sigma$ uncertainty.
Although  the LSF calibration as a function of \wl, channel, and aperture is not yet well determined (Kruk et al. 2001), we accounted for this
by considering two extreme assumptions (free vs. fixed LSF) and adopting an average column density that includes this systematic effect in the errors. 

 For Case 3 we used high resolution STIS E230H spectra to provide resolved \feii\, \wll2344, 2382 line profiles, the only \feii\ lines available in the existing archival high resolution spectra of \bd28. Profile fits to the STIS spectra alone found three components at $v_{hel}= -2.4, -9.7$, and $-17.0 \kms$ (listed in order of decreasing \feii\ column density) and a total \feii\ column density of $\log N$(\feii)$= 13.81 \cmsq$ (see Figure \ref{feiifit}).  We also performed a simultaneous fit of \fuse\ and STIS data using {\tt Owens.f}. This fit found the velocity separations of the three components to be 3.5 and 10.9 $\kms$ and $\log N$(\feii)$=13.79 \cmsq$, compared with 7.3 and 7.3 $\kms$ for the STIS data alone. These differences $(3-4 \kms)$ reflect the level of uncertainty of these fits. 

For the Case 3 fits, the relative ratios of
each component for the neutral species was very different from that in  \feii.  
Specifically,  the weakest (bluest) \feii\ component had no contribution from \di, \ni, or \oi.
This could be due to ionization effects or iron might be depleted in different ways in different
environments, while \ni\ and \oi\ are only lightly depleted (Savage \&
Sembach 1996). It appears  that the fit of the neutral species
suffered from the use of the \feii\ lines, showing a higher reduced
\chisq. We believe that the neutral species are not  fit well by the \feii\ velocity structure
because the $b$-values and/or the velocity structure appears to be
different. Since we do not resolve the component
structure with \fuse, we rely on the one component solution for \di,
\ni, and \oi.

The column density errors were calculated by
computing new fits for different fixed 
values of $N$ away from the best fit column density and deriving  $\Delta$\chisq\ as a function of $N$.  The $3\sigma$ error
is given by $\Delta$\chisq = 9.  The average gives
 log \ndi\ $= 14.99 \pm 0.03 \cmsq (2\sigma)$.

The COG and profile fitting analyses give the same \di\ column density and have error estimates that differ only slightly at the $2\sigma$ level. It is reassuring that both approaches 
give results within 5\% of each other.  The errors  from the COG analysis are more conservative than the ones coming from the profile fits. The reason for this is that moving the continuum by $\pm3\%$, as was done for the COG, we have probably overestimated the errors. Yet they seem to fairly reflect  the scatter  seen in the individual \eqw\ measurements.   Combining the results,  we adopted a straight average of \ndi\ from the COG and profile fitting analyses and arrive at the final \di\ column density log \ndi\ $= 14.99 \pm 0.05 \cmsq ~(2\sigma)$ for the \bd28\ line of sight.

  Figure \ref{oifits} compares the \oi\ fits for SiC2 spectra from the LWRS and MDRS observations. Combining the results from the different LSF
approaches (see Table \ref{dnotable}) gives log \noi$=16.21\pm 0.10 \cmsq~(2\sigma)$ from the profile fitting analysis.

Most of the \oi\ lines used in the profile fitting analysis are close to saturation which 
may make both this and the COG method sensitive to the velocity structure. We investigated this possibility by analyzing the effect of a three-component velocity structure on the COG analysis. The goal
of this investigation was to see if a one-component COG  could still
be applied with success on this line of sight. We computed a model for the line of sight for \di\ and \oi\ lines with three components as seen in the STIS \feii\ profiles.
Performing a one-component COG analysis of the model line profiles yielded \di\ and \oi\ column densities within 5\% of the model and an average effective
$b$ of 5.8 $\kms$. Although the velocity structure found in \feii\ is not
identical to that for the neutral species, it is the best available. 
For the \bd28\ line of sight   a one-component COG gives
reliable results. We find that the effective $b$ from the analysis of the three-component simulated profiles is 
consistent with the effective $b$ from the analysis of the actual data.
The final value adopted for the \oi\ column density is  log \noi $= 16.22\pm 0.10 \cmsq~(2\sigma)$.

  Examples of the fits to the \ni\,\wll952 lines are shown in Figure \ref{nifits} and the results listed in Table \ref{dnotable}. The mean fit gives log \nni $=15.55 \pm 0.13 ~(2 \sigma)$.

\section{RESULTS AND DISCUSSION}

The abundances derived in the previous sections are summarized in Table \ref{dhtable} along with their ratios.  The $2\sigma$ errors in the abundance ratios are those from the individual column densities combined in quadrature. We find that D/H $=(1.39\pm 0.21) \EE{-5}~(2\sigma)$ toward \bd28.  G\"olz et al. (1998) observed \bd28\ with the T\"ubingen echelle spectrograph ($\lambda/\Delta\lambda \sim$ 10,000) on the \orf\ mission and found D/H $=(0.8^{+0.7}_{-0.4}) \EE{-5} (1\sigma)$. Our result is at the upper end of their range.   The \fuse\ result for \bd28\ is just marginally consistent with the higher D/H value found for the more distant sight line toward Feige 110 (D/H $=(2.1\pm 0.70) \EE{-5}$) by Friedman et al. al. (2001).  The total neutral gas column toward \bd28\ is very similar to that toward \gvel, \zpup, and \dori\ (at distances of 250--500 pc), the sightlines studied with IMAPS to measure D/H (Jenkins et al. 1999, Sonneborn et al. 2000).  D/H toward \bd28\ is in the middle of the range of values found in the IMAPS studies.    

Our result for D/H toward \bd28\ is similar to values found on lines of sight in the Local Bubble by \fuse\ (Moos et al. 2001) and \hst\ (Linsky et al. 1998), even though \bd28\ lies just outside the Local Bubble. The total \hi\ column density toward \bd28\ is  log \nhi = 19.85, much higher than the other LISM sight lines studied thus far by \fuse\ for D/H (except for Feige 110). The \nai\ contours for the LISM (Sfeir et al. 1999) show that \bd28\ is located in a region of higher column density just beyond the edge of the Local Bubble, with much lower densities in the intervening $\sim100$ pc. The \hi\ column toward \bd28\ is large enough that the local cloud is a negligible contribution.  Our abundance measurements are therefore dominated by the gas $\sim100$ pc from the Sun.   As discussed by Moos et al. (2001), there is now considerable evidence for a constant D/H within the Local Bubble.  Beyond $\sim100$ pc, however, or more specifically outside the Local Bubble and log \nhi\ $ge 19.5$, there appears to be a considerable spread in D/H as larger amounts of gas are probed. 

The total \h2\ column density is $\sim 1.3 \EE{15} \cmsq$, and the molecular fraction $f({\rm H_2}) \sim 3.8 \EE{-5}$.  This low value of $f({\rm H_2})$ is typical of lower \hi\ column density ISM sight lines (Tumlinson et al. 2001).  The abundance  of \h2\ at $\sim100$ pc indicates that some molecular gas is present, but the low column densities and high excitation temperature are consistent with a low density environment where UV pumping is probably significant.

The profile fitting and COG analyses for \di\ and \oi\ consistently found an effective $b$-value of $\sim 5 \kms$.  Although \fuse\  lacks the spectral resolution to sepatate thermal and non-thermal effects on the neutral gas line profiles, we can set an upper limit to the temperature.  For \di\ $b=5.0 \kms$ corresponds to $T\le 6000$ K.  For \oi, however, the same $b$ corresponds to $T\le$40,000 K.  There must either be a significant non-thermal contribution to the line broadening of the neutral species, or the component structure on the \bd28\ line of sight has a significant effect of the line shapes.  Very high resolution spectra of species such as \nai\ are needed to determine the component structure, $T$ and $v_{turb}$ of the neutral gas as a function of velocity.

The O/H ratio toward \bd28\ is $(2.37\pm 0.55) \EE{-4} ~(2\sigma)$, a value that is anomalously low with respect to the Local Buble and the ISM.  The \bd28\ O/H is marginally consistent with the mean O/H ratio  in the ISM ($<$O/H$>_{ISM}$) of $(3.19\pm0.28) \EE{-4} ~(2\sigma)$ found by Meyer et al. (1998) in a  study of \oi\ \wl1355 on 13 sight lines toward OB stars at  distances of 130 -- 1500 pc.  Recently, however, Meyer (2001) updated $<$O/H$>_{ISM}$ to $(3.43\pm0.28) \EE{-4}$ as a result on a revised $f$-value for the \wl1355 transition.   The \fuse\ O/H ratio toward \bd28\ is well outside the estimated uncertainties in the revised $<$O/H$>_{ISM}$.  In the Local Bubble $<$O/H$>_{LB}=(3.9\pm0.3) \EE{-4}~(1\sigma)$ (Moos et al. 2001),  slightly higher than Meyer's revised value for the ISM, although consistent within their respective uncertainties.  Moos et al. (2001) also finds the same value of the mean O/H when the more distant sight line to Feige 110 is included. Of the five sight lines for which O/H is measured by \fuse, only \bd28\ gives a value that disagrees with $<$O/H$>_{ISM}$ and $<$O/H$>_{LB}$. 
If correct, the gas just outside the Local Bubble has a low O abundance.   This implies that the gas toward \bd28\ beyond the Local Bubble would appear to have a different chemical history than that sampled in the Meyer et al. study.   

We have searched for other possible explainations for the low O/H ration toward \bd28.    \oi\ is not vulnerable to ionization effects (Jenkins et al. 2000), so the \oi-to-\hi\ ratio should accurately reflect the total O/H abundance ratio.  If the error is in the \hi\ column density, then correcting O/H to obtain agreement with $<$O/H$>_{ISM}$ would result in a large D/H ratio ($\sim2.0 \EE{-4}$). Such a D/H value for \bd28\ would be similar to that found for Feige 110 (Friedman et al. (2001) and \gvel\ (Sonneborn et al. 2000).  However, a $\sim$40\% increase in \nhi\ would be inconsistent with the available spectra.

The D/O ratio for \bd28\ appears to be high (Moos et al. 2001, H\'ebrard et al. 2001), arguing for \noi\ as the anomalous quantity. Agreement with $<$O/H$>_{ISM}$ would require a $\sim40$\% increase in \noi.  One or more of the \oi\ lines analyzed might actually be saturated, even though our analysis shows (Fig. \ref{oifits}) that the three \oi\ lines are unsaturated.  Line saturation, possibly as the result of narrow (cold gas) components  would result in an underestimated column density. The effect of potential line saturation in \noi\ was examined by excluding the strongest \oi\ line (\wl930) from the fitting analysis, leaving only the two weaker \oi\ lines.  For this case we found log \noi\ = 16.28, which is within $2\sigma$  of the best fit.  We cannot determine whether this difference in $N$ is the result of saturation, fixed pattern noise, LSF uncertainties, component structure, or some other cause. If this were the correct \oi\ column density, the O/H ratio would be $2.72 \EE{-4}$, still well below $<$O/H$>_{ISM}$ and $<$O/H$>_{LB}$.

The N/H ratio toward \bd28\ is $(5.08\pm 1.66) \EE{-5} ~(2\sigma)$, a value that is marginally consistent with the mean ISM value of N/H $=(7.5\pm0.8) \EE{-5} ~(2\sigma)$ found for six OB stars by Meyer et al. (1997).  However, the \nii\ column density could be significant (see below).  If so, then the N/H ratio toward \bd28\ could well be consistent with the mean ISM result.  

In the local ISM, the abundance of \ni\ has been found to be reduced by ionization effects on some sight lines (Jenkins et al. 2000, Moos et al. 2001). \ni\ has a photoionization cross section larger than that of \hi\ and a large fraction of it can be ionized.
There are only two \nii\ transitions in the \fuse\ bandpass (\wl915.6 and \wl1085.5) and
both of them are heavily saturated on all but the lowest column density sight lines. We derived 
a lower limit of log \nnii\ $>14.2$ using both optical depth technique
and profile fitting. Since we know that some \nii\ is
associated with the neutral component, we estimate an upper limit
assuming that the  $b$-value for \nii\  is greater than that of the
neutral gas. These assumptions give an upper limit of log \nnii\ $<15.7$. We cannot rule out the possibility that an apreciable fraction of the nitrogen on the sight line toward \bd28\ could be singly ionized.

Objects like the hot subdwarf \bd28\ are important targets for studying \di\ and \oi\ in the ISM with \fuse\ because they can sample regions of space beyond that accessible with white dwarfs ($d < 100$ pc) and closer that that sampled by the lightly-reddened O stars observable by \fuse\ ($d>1$ kpc).  Most of the nearer O stars far exceed the \fuse\ brightness limit.

\begin{figure}
\plotone{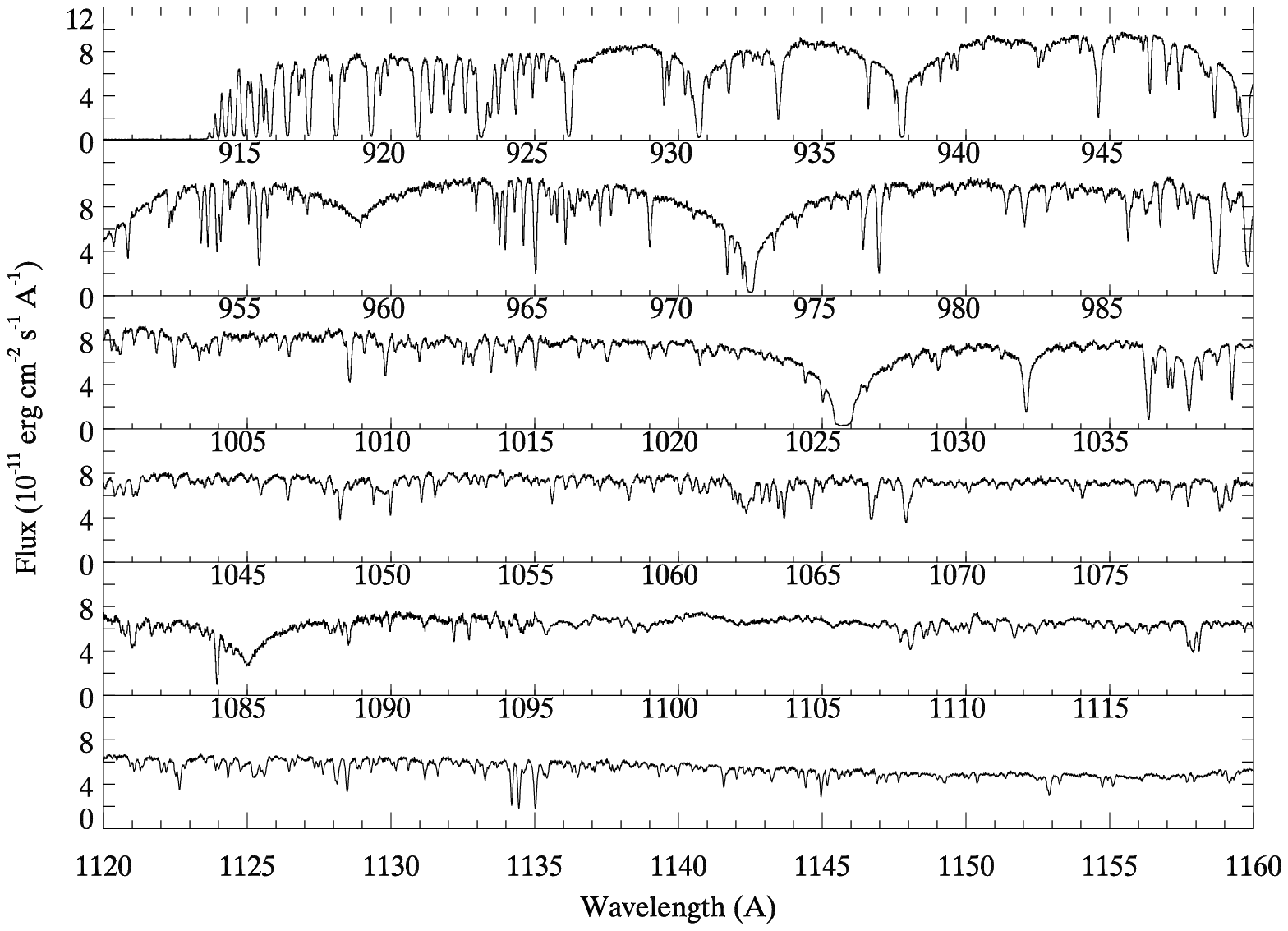}
\caption{FUSE co-added LWRS spectra of \bd28\ obtained on 2000 July 16. The spectral region 910-990 \AA\ is from the SiC1 channel, 1000-1080 \AA\ from LiF1, 1080-1095 \AA\ from SiC2, and 1095-1160 \AA\ from LiF1.  The data have been binned by a factor of 2 for display purposes. \label{bdspec}}
\end{figure}

\begin{figure}
\epsscale{0.75}
\plotone{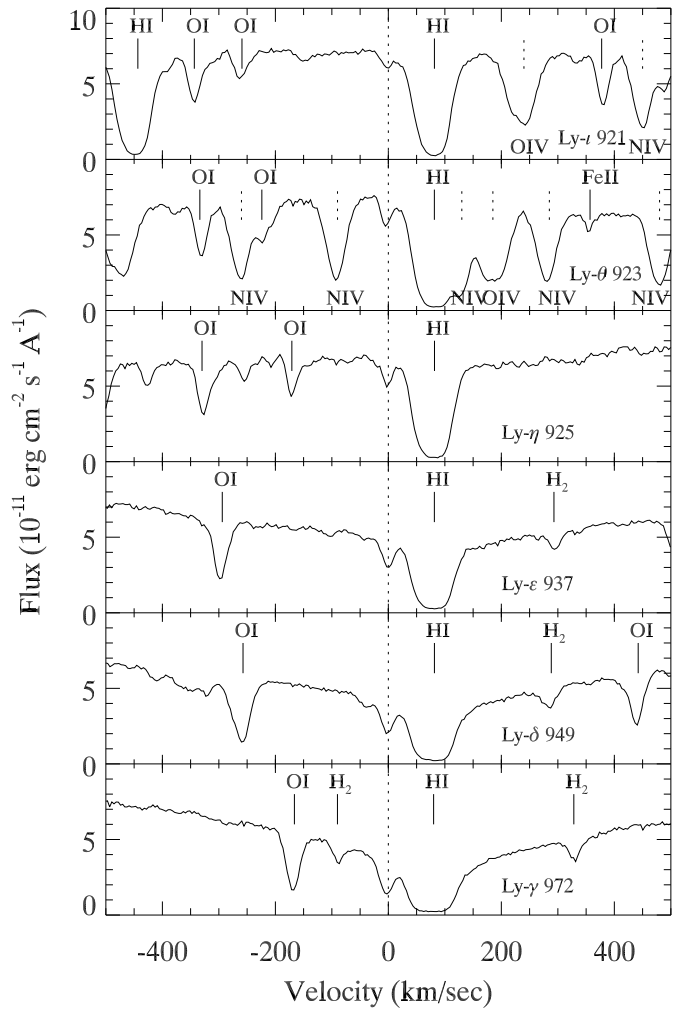}
\caption{The higher Lyman lines in \bd28\ from the SiC1 LWRS spectra are shown on a velocity scale.  The zero point is  set to the observed center of the \di\ feature.  The principal interstellar atomic transitions are indicated above the spectrum and stellar transitions below.  \label{lyspec}}
\end{figure}

\begin{figure}
\plotone{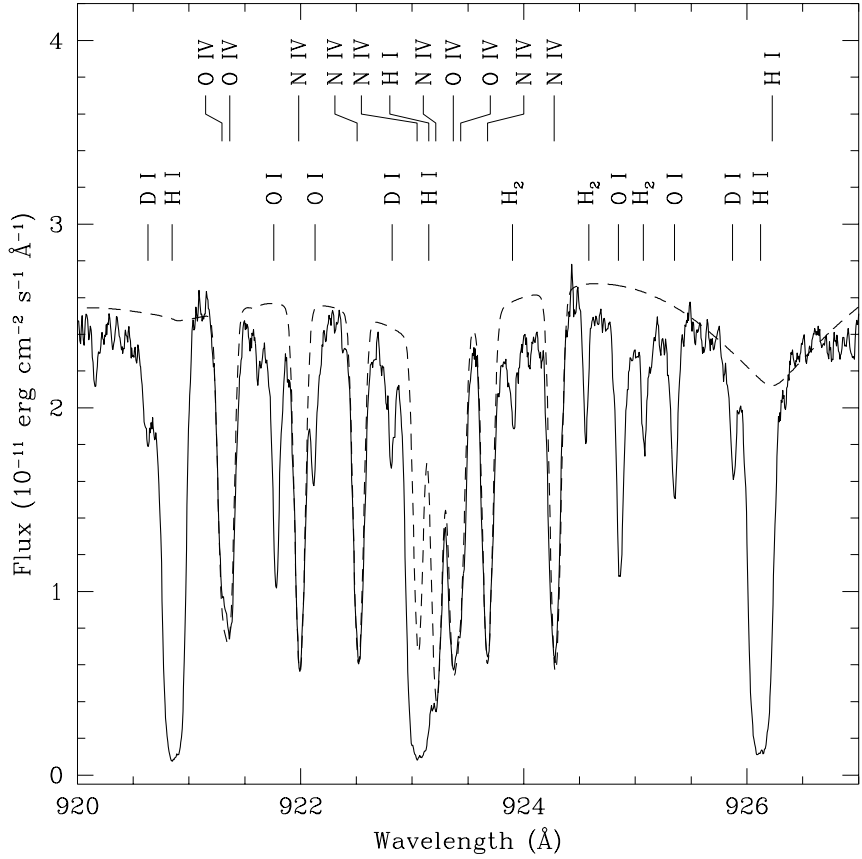}
\caption{Comparison between a portion of the \fuse\ spectrum (solid line) and a synthetic spectrum (dashed line) computed
with $T_{\rm{eff}} = 82,000$ K, $\log({\rm{He/H}}) = -1.0$, $\log g =
6.2$, and incorporating solar nitrogen and oxygen abundances.  The
photospheric lines are labelled at the top of the figure. The ISM
lines are labelled just above the spectrum.\label{model923}}
\end{figure}

\begin{figure}
\plotone{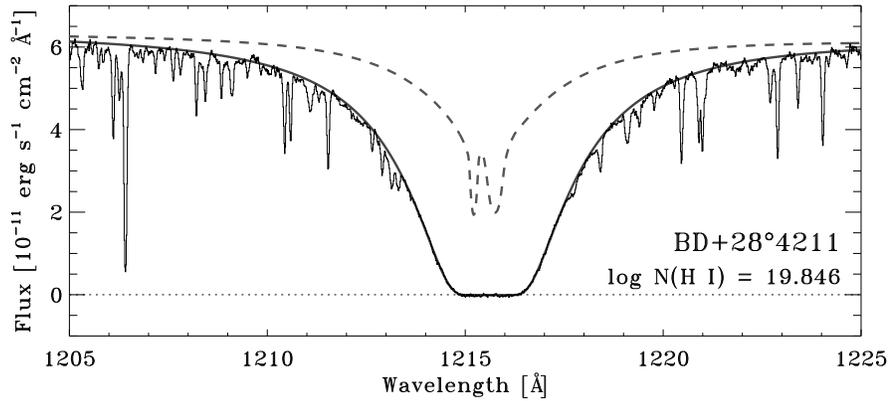}
\caption{The wavelength region surrounding interstellar \lya\ towards \bd28.  
The histogram shows the coadded STIS E140M observations of \bd28.  The
dashed gray line shows the adopted stellar model ($T_{eff} = 82,000$
K, $\log g = 6.2$, and $\log {\rm He/H} = -1.0$) modified by a second
order Legendre polynomial to bring it into agreement with the observed
flux distribution. The two narrow absorption components in the stellar profile are the cores of \hi\ and \heii. The solid line shows the best-fit \hi\ model times
this stellar continuum.  The best-fit interstellar parameters are
$v_{hel} = -13.3\pm0.8$ km s$^{-1}$ and $\log N(\mbox{\hi}) =
19.846^{+0.034}_{-0.036}$ (both $2\sigma$ uncertainties).
\label{lyalpha}}
\end{figure}

\begin{figure}
\plotone{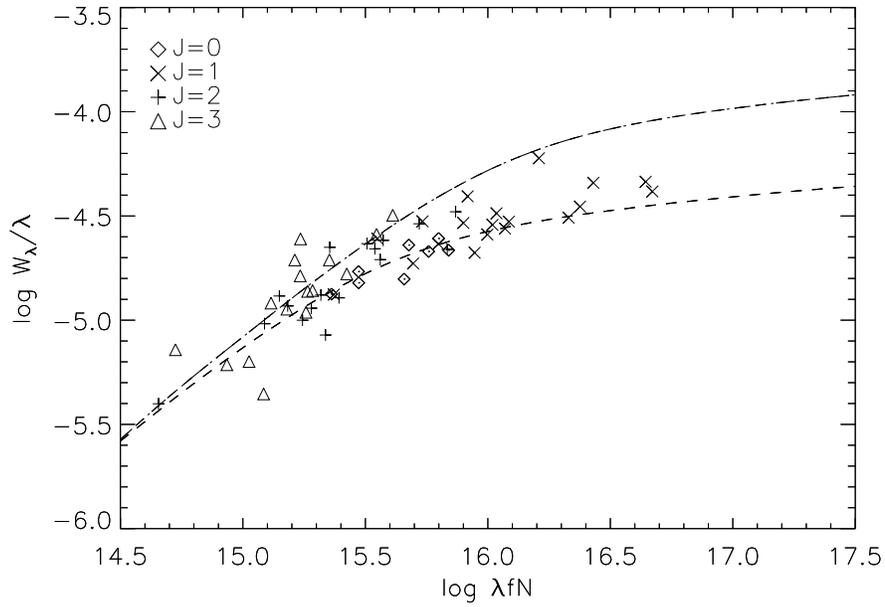}
\caption{Curve of growth for \h2\ with fits corresponding to $b=2.8 \kms$
(dashed line) and $b=8.5 \kms$ (dot-dashed line), which were the minimum and
maximum $b$ values found in the fit of the individual rotational levels. The $1\sigma$ error bars are comparable to the scatter in the \eqw\ measurements.  They were omitted from the figure for clarity.\label{h2cog}}
\end{figure}

\begin{figure}
\epsscale{0.80}
\plotone{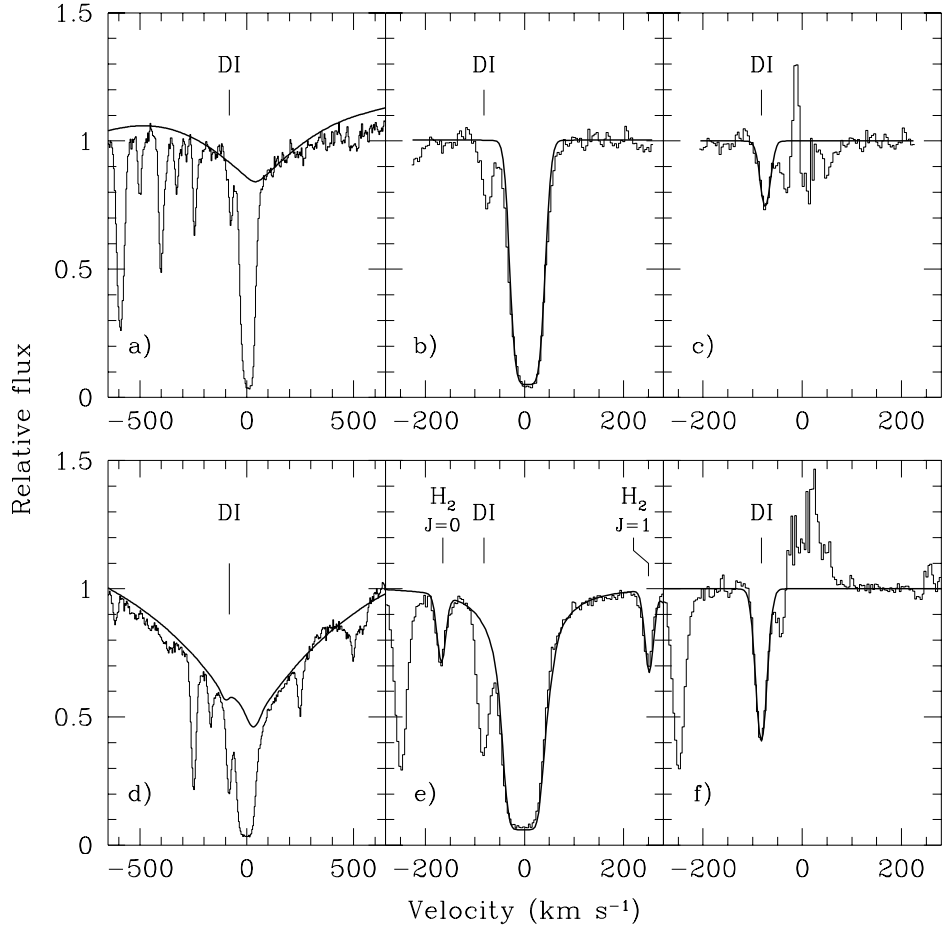}
\caption{Continuum normalization for \di\ \eqw\ measurements. Panels a) - c) illustrate the method used to measure \eqw\ of \di\  \wl 925.974 (\lyeta) in the SiC2 channel and LWRS aperture. Panel a) shows the stellar model placement (smooth, solid line)  over the data (histogram-style line). Panel b) shows the ISM model (\hi\ + \h2\ only) relative to the spectrum normalized by the stellar model. Panel c) shows the ISM-normalized spectrum from b) and the Gaussian fit to the \di\ \lyeta\ profile used to measure the area of the line under the continuum. The red side of the reconstructed \di\ profile is affected by large residuals from dividing by the near-zero flux in the core of \hi. Panels d)-f) show the same method applied to \di\ \lyg\  in the same channel and aperture. The weak feature in the blue wing of the stellar \lyg\ model profile (panel d) is \heii, while the stronger feature directly below it in the observed spectrum is interstellar \di. \label{cogmethod}}
\end{figure}

\begin{figure}
\plotone{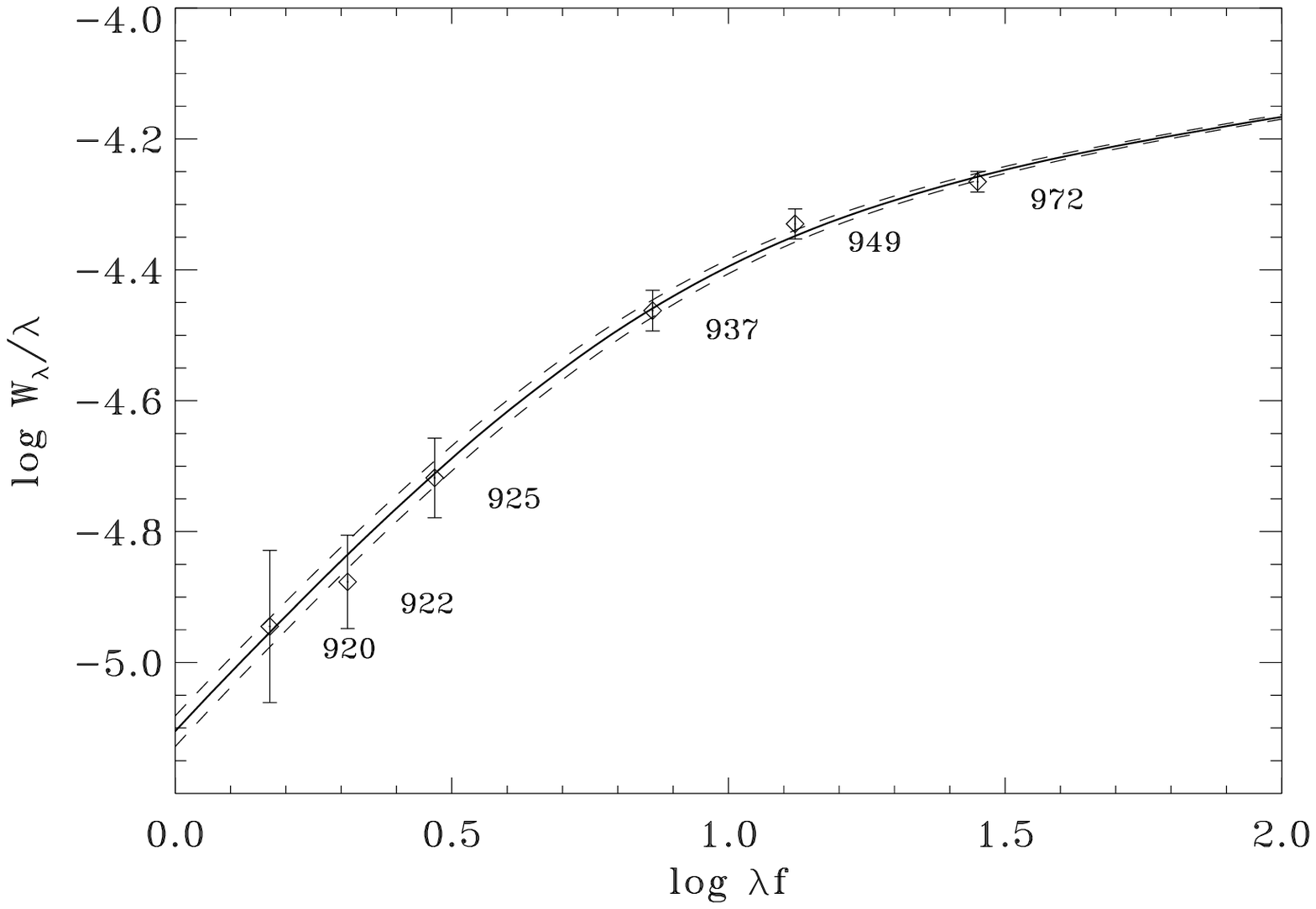}
\caption{The best fit curve of growth for \di\ (solid line) corresponding to log \ndi\ $=14.99\pm 0.10~(2\sigma)$ and $b=5.2 \kms$. The dashed lines are the $1\sigma$ errors in the fit. The equivalent width data points,
shown with the wavelength of the transition and $1\sigma$ error bars, are the weighted average of \eqw\ from the SiC1 and SiC2 channels and the LWRS and MDRS observations.\label{diallcog}}
\end{figure}

\begin{figure}
\plotone{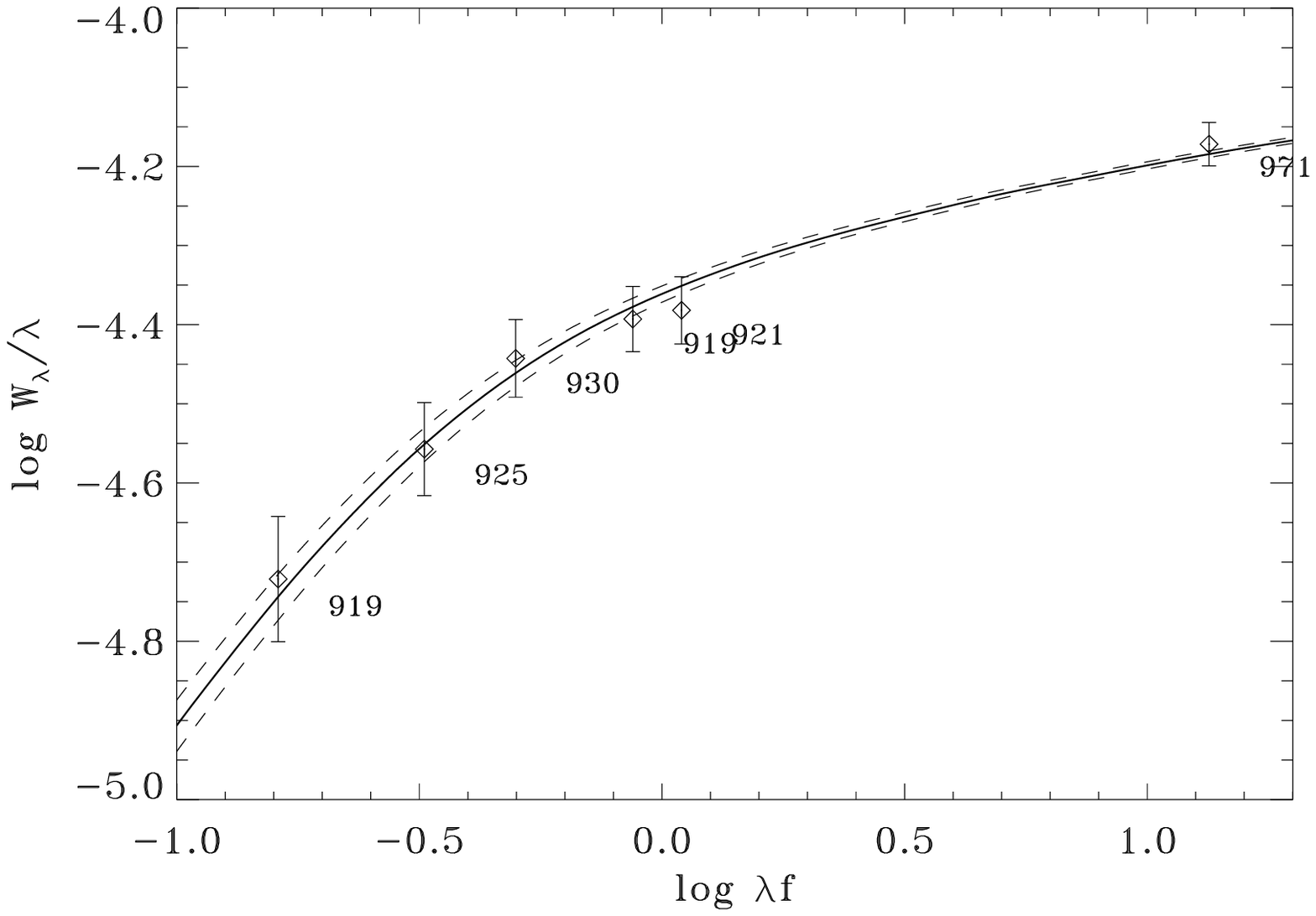}
\caption{The best fit curve of growth for \oi\ (solid line) corresponding to log \noi\ $=16.23\pm 0.08 (2\sigma)$ and $b=4.46 \kms$. The dashed lines are the $1\sigma$ errors in the fit. The equivalent width data points,
shown with the wavelength of the transition and $1\sigma$ error bars, are the weighted average of \eqw\ from the SiC1 and SiC2 channels and the LWRS and MDRS apertures.\label{oiallcog}}
\end{figure}

\begin{figure}
\plotone{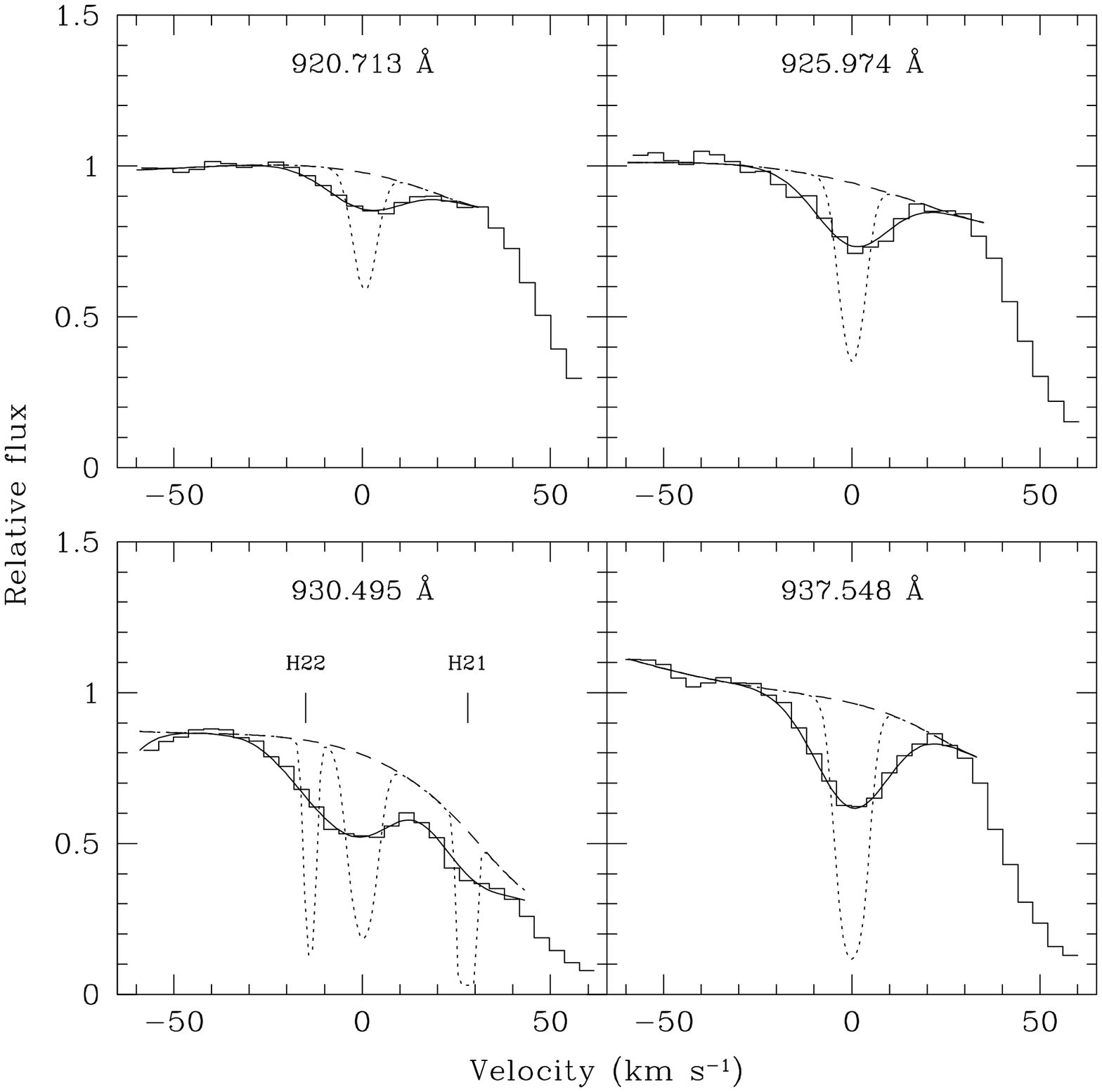}
\caption{ The fits of \di\ \lye\,\wl937.5, \lyz\,\wl930.5, \lyeta\,\wl926.0, and \lyi\,\wl920.7 for the  SiC2 channel and LWRS aperture are shown on a relative velocity scale. The histogram lines represent the data, the thick solid lines are the final fits to the spectrum, and the dashed lines are the fits after deconvolution with the LSF. Note that no \di\ lines stronger than \lye\ were used since this transition is already optically thick ($\tau_0 \sim 2$). Note also \di\ \lyz\ is  blended with \h2$(J=1)$ and \h2$(J=2)$.  The profile fitting technique recovers the information for the \di\ profile contained in the data. \label{difits}}
\end{figure}

\begin{figure}
\plotone{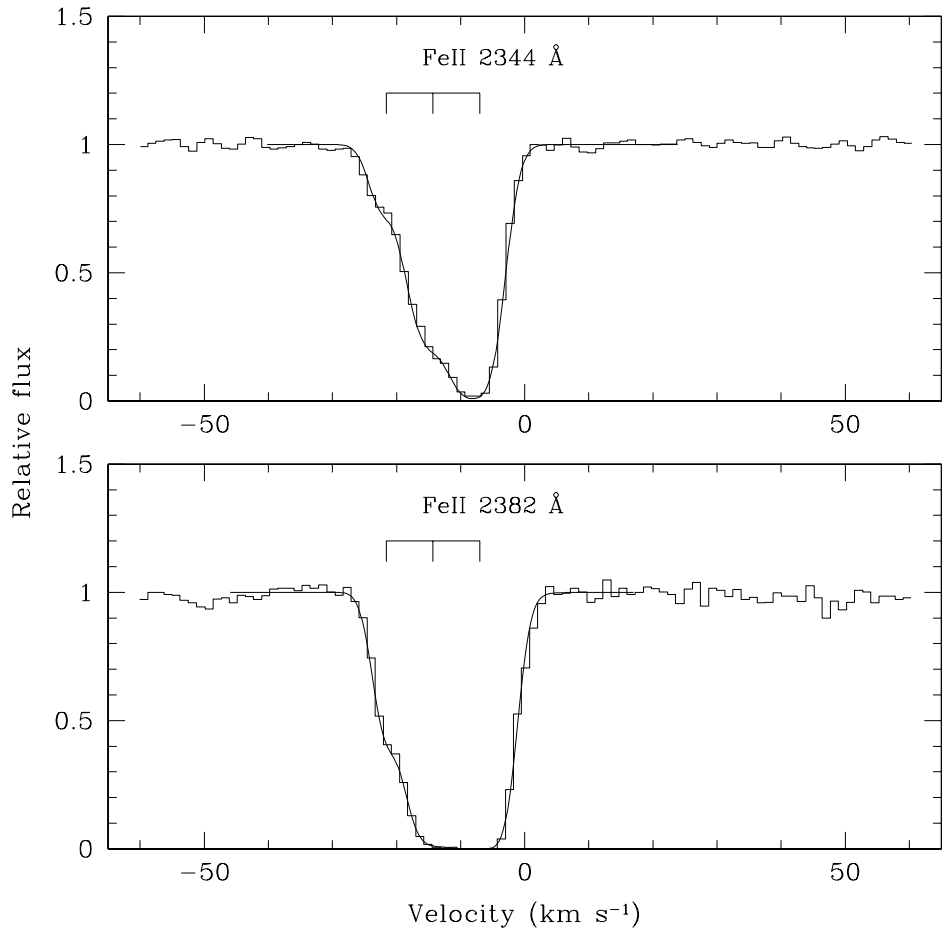}
\caption{\feii\,\wl2344 and \wl 2382 line profiles from \hst/STIS E230H spectra of \bd28\ are compared with the best fit model and shown on a heliocentric velocity scale.  The smooth solid line is the best fit obtained with three components.  The histogram line is the normalized spectrum.  70\% of the total column density is in the dominant component (right), 27\% in the middle component, and 3\% in the weakest component (left). The two stronger components are separeted by less than $8 \kms$ and are not resolved in \feii\ features present in the \fuse\ spectra of \bd28.\label{feiifit}}
\end{figure}

\begin{figure}
\plotone{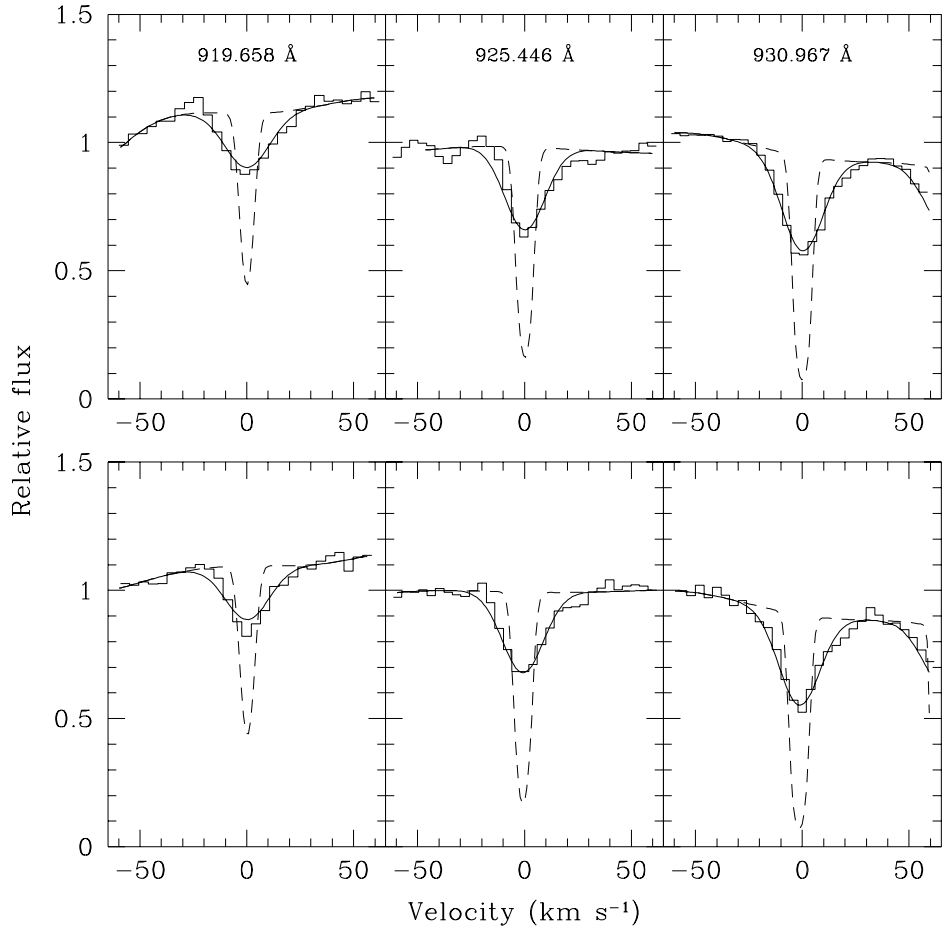}
\caption{Profile fits to three  optically thin \oi\ lines (\wll 919.92, 925.45, and 930.26). The top panels shows the fits for the  SiC2 channel LWRS aperture spectra. The bottom panel shows 
the same transitions for the same channel but MDRS aperture. The definitions of the different line styles are the same as Figure \ref{difits}. Both data 
sets are very similar, with a LSF $FWHM \sim 20 \kms$ for LWRS and  
$\sim 18 \kms$ for MDRS.
\label{oifits}}
\end{figure}

\begin{figure}
\plotone{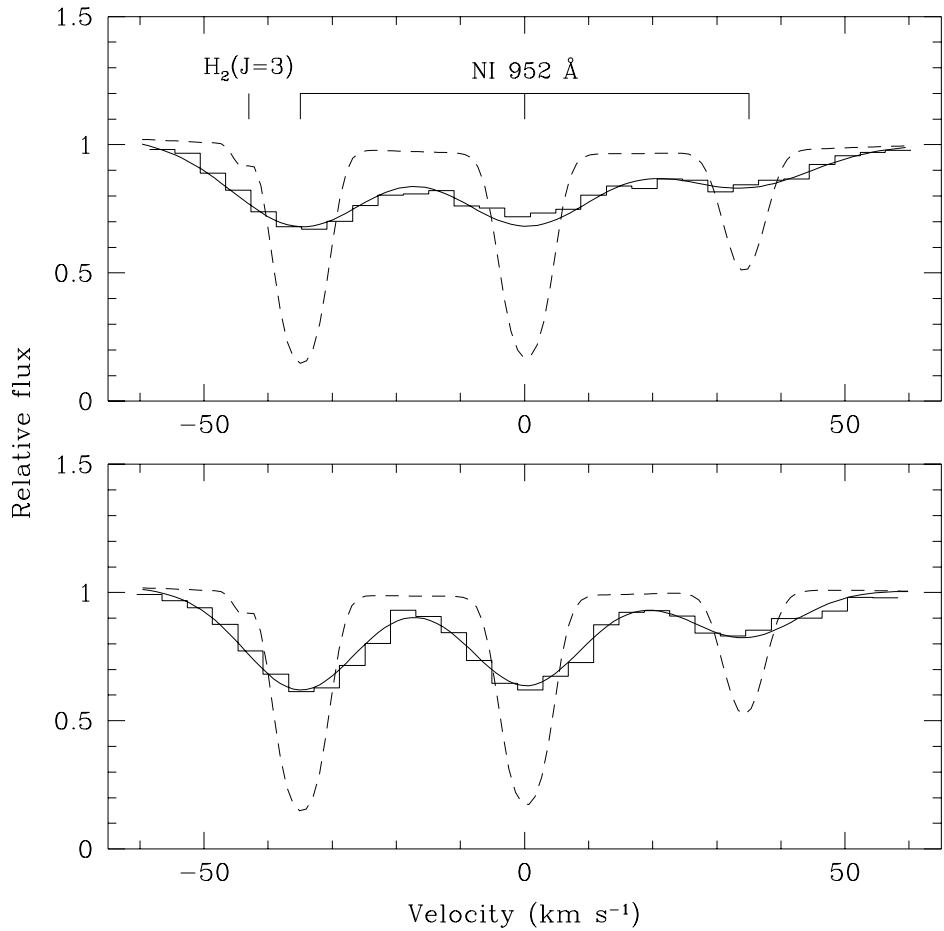}
\caption{Profile fits to the \ni\,\wl952 multiplet. The top panels shows the fits to the spectrum for the SiC2 channel and the LWRS aperture. The bottom panel is the same channel for the MDRS aperture. The definitions of the different line styles are the same as Figure \ref{difits}. The spectral resolution in the two data sets are very similar. Note the \h2\ L$14-0$ P(3) line  on the blue side of the triplet. This \ni\ triplet is the only one available in the \fuse\ bandpass that is not optically thick and free of significant interference by  lines of other species.\label{nifits}}
\end{figure}

High-quality spectra of the type shown in Figures \ref{bdspec} and \ref{lyspec} are needed to understand these objects, and the observational techniques to obtain them with \fuse\ are now available. Over time issues like stellar continuum placement and stellar line identifications may be better understood, allowing even more precise interstellar abundance measurements.

\acknowledgements

This work is based on data obtained for the Guaranteed Time Team by the NASA-CNES-CSA FUSE mission operated by the Johns Hopkins University. Financial support to U. S. participants has been provided in part by NASA contract NAS5-32985 to Johns Hopkins University.  Support for French participation in this study has been provided by CNES.

\clearpage
\onecolumn

\clearpage

\clearpage

\clearpage

\clearpage

\clearpage

\clearpage

\clearpage

\clearpage

\clearpage

\clearpage

\clearpage

\begin{deluxetable}{llcccrcc}
\tablewidth{0pt}
\tablecaption{{\it FUSE} Observations of \bd28 \label{fpsplit}}
\tablehead{
\colhead{Program ID} & \colhead{Date} & \colhead{Aperture} &  
\colhead{FP-split}& \colhead{Offset} & $\Delta\lambda$ & \colhead{Exposures} & 
\colhead{Exp. Time}  \\ 
\colhead{} & \colhead{} & \colhead{} &  \colhead{pos No.} &\colhead{(pixels)} & 
(\AA) & \colhead{} & \colhead{(s)} 
}
\startdata
M1080901 & Jun 13 2000 & LWRS & \nodata & \nodata & \nodata & 1--4 & 2192 \\
M1040101 & Jul 16 2000 & LWRS & 1&  $-27$ & $-0.17$ & 1--14 & 6251   \\
         & &  & 2 & 0 & 0.00 & 14--25  & 5073   \\
         & &  & 3 & +7 & +0.04 & 26--30 & 2454  \\
         & &  & 4 & +36 & +0.23 & 31--36  & 2799  \\
M1040105 & Sep 19 2000 & LWRS & 3 & +7 & +0.04 & 1--8 & 3691 \\
         & &  &  4 & +36 & +0.23 & 9--17 & 4146 \\ 
M1040102 & Jul 17 2000 & MDRS &  1 & $-12$ & $-0.08$ & 1--13 & 6279 \\
         & &  & 2 & 0     & 0.00 & 14--30 & 8211 \\
         & &  & 3 & +5    & +0.03 & 31--42 & 5795  \\
         & &  & 4 & +12   & +0.08 & 43--52  & 4540  \\       
\enddata
\end{deluxetable}

\begin{deluxetable}{lcc}
\tablecolumns{8}
\tablewidth{0pt}
\tablecaption{Properties of BD\,+28\degr\,4211}
\tablehead{
\colhead{Quantity} &
\colhead{Value} &
\colhead{Reference}
}
\startdata
 Spectral Type & sdO & 1 \\
 $l$ & $81.87\degr$ & 2 \\
 $b$ & $-19.29\degr$ & 2 \\
 $d$\tablenotemark{a} \ (pc)& $104\pm 18$ & 2 \\
 $V$ & 10.53 & 3 \\
$U-B$ & $-1.26$ & 3 \\
$B-V$ & $-0.34$ & 3\\
$v_{hel}$ & $+20.2 \pm 0.8 \kms$ & 4 \\
$T_{\rm eff}$\ (K) & 82,000 & 1 \\
$\log g$\ (cm s$^{-2}$) & 6.2 & 1 \\
$\log ({\rm He / H})$ & $-1.0$ & 1 \\
\enddata
\tablenotetext{a}{Trigonometric parallax}
\tablerefs{(1) Napiwotzki 1993; (2)\ Perryman et al. 1997;
(3)\ Kidder et al.\ 1991; (4) This paper (\S \ref{nhi}).}
\end{deluxetable}

\begin{deluxetable}{ccl}
\tablewidth{0pt}
\tablecaption{\h2\ Column Densities\tablenotemark{a}\label{nh2table}}
\tablehead{
\colhead{$J$} & \colhead{No. Lines} & \colhead{log(\nh2)} 
}
\startdata
0 & 4 & $14.57^{+0.45}_{-0.15}$  \\
1 & 8 & $14.76^{+0.16}_{-0.06}$  \\
2 & 13 & $14.39^{+0.07}_{-0.04}$  \\
3 & 14 & $14.17^{+0.05}_{-0.03}$  \\
4 & 3 & $<13.68$ \\
\enddata
\tablenotetext{a}{Errors are $2\sigma$. Value for $J=4$ is a $3\sigma$ upper limit.}
\end{deluxetable}

\begin{deluxetable}{ccccccc}
\tablewidth{0pc}
\tablecaption{\di and \oi\ Equivalent Widths\tablenotemark{a}\label{doeqw}}
\tablehead{Line & $\lambda$ & $\log f\lambda$ & \eqw\ (LWRS) & ~ & \eqw\ (MDRS) & $\langle$\eqw$\rangle$ \\
~ & (\AA) & ~ &  (m\AA) & ~ &  (m\AA) & (m\AA) }
\startdata
~ & \di\ & ~ & SiC1 ~~~~ SiC2 & ~~ & SiC1 ~~~~ SiC2 & ~ \\
Ly$\iota$ & 920.713 & 0.17 & ~~~\nodata ~~~~~ $10.4\pm4.7$ & ~~~ & ~~~\nodata ~~~~~ $10.5\pm4.4$ & $10.5\pm3.2$ \\
Ly$\kappa$ & 922.899 & 0.31 & $14.1\pm6.0$ ~~ $13.7\pm3.8$ & ~~~ & $10.9\pm3.0$ ~~~~ \nodata ~~~ & $12.3\pm2.2$ \\
Ly$\eta$ & 925.974 & 0.47 & $18.5\pm7.5$ ~~ $18.3\pm7.0$ & ~~~ & $19.8\pm5.0$ ~~ $16.0\pm4.0$ & $17.7\pm2.7$ \\
Ly$\epsilon$ & 937.548 & 0.86 & $34.4\pm5.0$ ~~ $30.3\pm5.7$ & ~~~ & $34.1\pm4.0$ ~~ $29.1\pm5.0$ & $32.3\pm2.4$ \\
Ly$\delta$ & 949.485 & 1.12 & $42.0\pm4.6$ ~~ $43.8\pm4.4$ & ~~~ & $48.5\pm5.0$ ~~ $44.0\pm5.5$ & $44.4\pm2.4$ \\
Ly$\gamma$ & 972.272 & 1.45 & $50.6\pm4.0$ ~~ $55.0\pm4.4$ & ~~~ & $52.3\pm3.0$ ~~ $54.6\pm5.0$ & $52.8\pm1.9$ \\
~ & ~ & ~  & ~~~ &  ~ \\
~ & \oi\ & ~ & SiC1 ~~~~ SiC2 & ~~~ & SiC1 ~~~~ SiC2 & ~ \\
~ & 919.917\tablenotemark{b} & -0.79 & $15.6\pm 1.2$ ~~ $20.8\pm 1.2$ & ~~~ & $16.7\pm 1.3$ ~~$16.6\pm 1.3$ & $17.5\pm0.6$ \\
~ & 925.446\tablenotemark{b} & -0.49 & $27.4\pm 1.3$ ~~ $23.7\pm 1.3$ & ~~~ & $28.1\pm 1.4$ ~~$23.1\pm 1.3$ & $25.7\pm0.7$ \\
~ & 930.257\tablenotemark{b} & -0.30 & $32.5\pm 1.4$ ~~ $34.0\pm 1.4$ & ~~~ & $34.8\pm 1.5$ ~~$33.1\pm 1.4$ & $33.6\pm0.7$ \\
~ & 919.658\tablenotemark{c} & -0.06 & $34.5\pm 1.3$ ~~ $41.5\pm 1.4$ & ~~~ & $43.1\pm 1.4$ ~~$30.9\pm 1.3$ & $37.2\pm0.7$ \\
~ & 921.857\tablenotemark{c} & 0.04 & $41.7\pm 1.4$ ~~ $37.6\pm 1.4$ & ~~~ & $41.3\pm 1.5$ ~~$33.4\pm 1.4$ & $38.3\pm0.7$ \\
~ & 971.738\tablenotemark{c} & 1.13 & $64.5\pm 1.5$ ~~ $65.3\pm 1.5$ & ~~~ & $68.6\pm 1.6$ ~~$63.6\pm 1.5$ & $65.4\pm0.8$ \\
\enddata
\tablenotetext{a} {Errors  in this table are $1\sigma$.}
\tablenotetext{b} {Singlet. Used $f$-values from Morton (1991).}
\tablenotetext{c} {Triplet. Used sum of $f$-values from Morton (2000, private communication).}
\end{deluxetable}
\clearpage

\begin{deluxetable}{cccc}
\tablewidth{0pt}
\tablecaption{\noi from Curve of Growth\tablenotemark{a}\label{oitable}}
\tablehead{Channel & Aperture &  log($N$(OI))  & $b$  \\
~ & ~ & $(\cmsq)$ & $(\kms)$
 \ }
\startdata
SiC1 & LWRS   & $16.22 \pm 0.10$  & 4.46 \\
SiC1 & MDRS   &  $16.25 \pm 0.06$  & 4.83 \\
SiC2 & LWRS   & $16.29 \pm 0.16$  & 4.41 \\
SiC2 & MDRS  & $16.18 \pm 0.18$  & 4.24 \\
\enddata
\tablenotetext{a} {Errors in this table are $2\sigma$.}
\end{deluxetable}

\begin{deluxetable}{cccccccc}
\tablewidth{0pt}
\tablecaption{\di, \ni, and \oi\  Profile 
Fitting Results\tablenotemark{a}\label{dnotable}}
\tablehead{Case & No. & DOF & $\chi^2_{min}$ & $b$  & log \ndi\  & log \nni\  & log \noi\ \\
~ & Windows & ~ & ~ & ($\kms$) & $(\cmsq)$ &  $(\cmsq)$ &  $(\cmsq)$  }
\startdata
Case 1\tablenotemark{b} & 39 & 1444 & 1674 & 4.0 & $14.99\pm 0.05$  & $15.51\pm 0.05$  & $16.22\pm 0.09$ \\
Case 2\tablenotemark{c} & 39 & 1405 & 1441 & 4.2 & $14.99\pm 0.07$  & $15.57\pm 0.09$  & $16.22\pm 0.16$ \\
Case 3\tablenotemark{d} & 41 & 1527 & 1929 & 3.8 + 3.3 & $14.96\pm 0.05$  & $15.49\pm 0.05$  & $16.17\pm 0.06$ \\
\enddata
\tablenotetext{a} {Errors  in this table are $2\sigma$.}
\tablenotetext{b} {One Gaussian component, LSF fixed for each channel (SiC1 
or SiC2).}
\tablenotetext{c} {One Gaussian component, LSF free to vary from window to 
window.}
\tablenotetext{d} {Three Gaussian components as defined by \feii\ velocity 
structure, LSF fixed for each channel (SiC1 or SiC2). See text.}
\end{deluxetable}

\begin{deluxetable}{cccc}
\tablewidth{0pt}
\tablecaption{\di, \ni, and \oi\ Column Densities \tablenotemark{a}\label{dnobestfit}}
\tablehead{Ion & Profile Fitting  & Mean COG & Adopted  \\
~ & ~  & ~ & Solution \ }
\startdata
log(\ndi) $(\cmsq)$ & $14.99\pm0.03$  & $14.99\pm0.10$ & $14.99\pm0.05$ \\
log(\nni) $(\cmsq)$ & $15.55\pm0.13$  & \nodata & $15.55\pm0.13$ \\
log(\noi) $(\cmsq)$ & $16.21\pm0.10$  & $16.23\pm0.08$ & $16.22\pm0.10$ \\
\enddata
\tablenotetext{a} {Errors  in this table are $2\sigma$.}
\end{deluxetable}

\begin{deluxetable}{cc}
\tablewidth{0pt}
\tablecaption{Interstellar Abundance Toward \bd28\tablenotemark{a}\label{dhtable}}
\tablehead{Quantity & Value}
\startdata
log \ndi\  & $14.99\pm0.05 \cmsq $  \\
log \nni\  & $15.55\pm0.13 \cmsq $  \\
log \noi\  &  $16.22\pm0.10 \cmsq $  \\
log \nhi\  & $19.846^{+0.036}_{-0.034} \cmsq $  \\
D/H        &  $1.39 \pm 0.21  \EE{-5}$ \\
N/H        &  $5.08 \pm 1.66  \EE{-5}$ \\
O/H        &  $2.37 \pm 0.55  \EE{-4}$ \\
D/N        &  $0.274 \pm 0.092$ \\
D/O        &  $0.0590 \pm 0.0133$ \\
\enddata
\tablenotetext{a} {Errors in this table are $2\sigma$.}
\end{deluxetable}

\end{document}